\documentclass[trackchanges]{aastex7}
\usepackage{graphicx}	
\usepackage{amsmath}	
\usepackage{mathtools}
\usepackage{lineno}
\usepackage{threeparttable}

\begin{document}

\title[FRB Populations Through Galaxy Clusters Lenses]{Forecasting the FRB Population Observed Through Galaxy Cluster Lenses}

\author[orcid=0000-0002-4623-5329,sname=Sammons, gname=Mawson]{Mawson W. Sammons}
\affiliation{Department of Physics, McGill University, 3600 rue University, Montr\'eal, QC H3A 2T8, Canada}
\affiliation{Trottier Space Institute, McGill University, 3550 rue University, Montr\'eal, QC H3A 2A7, Canada}
\email[show]{mawson.sammons@mcgill.ca} 

\author[sname=Davies-Velie, gname=Evan]{Evan Davies-Velie} 
\affiliation{Department of Physics, McGill University, 3600 rue University, Montr\'eal, QC H3A 2T8, Canada}
\affiliation{Trottier Space Institute, McGill University, 3550 rue University, Montr\'eal, QC H3A 2A7, Canada}
\email{evan.davies-velie@mail.mcgill.ca}

\author[orcid=0000-0001-7166-6422, sname=Dobbs, gname=Matt]{Matt Dobbs} 
\affiliation{Department of Physics, McGill University, 3600 rue University, Montr\'eal, QC H3A 2T8, Canada}
\affiliation{Trottier Space Institute, McGill University, 3550 rue University, Montr\'eal, QC H3A 2A7, Canada}
\email{matt.dobbs@mcgill.ca}

\author[orcid=0000-0003-2739-5869, sname=Kader, gname=Zarif]{Zarif Kader}
\affiliation{Department of Physics, McGill University, 3600 rue University, Montr\'eal, QC H3A 2T8, Canada}
\affiliation{Trottier Space Institute, McGill University, 3550 rue University, Montr\'eal, QC H3A 2A7, Canada}
\email{zarif.kader@mail.mcgill.ca}

\author[sname=Siegel, gname=Seth]{Seth R. Siegel}
\affiliation{Perimeter Institute for Theoretical Physics, 31 Caroline Street N, Waterloo, ON N25 2YL, Canada}
\affiliation{Department of Physics, McGill University, 3600 rue University, Montr\'eal, QC H3A 2T8, Canada}
\affiliation{Trottier Space Institute, McGill University, 3550 rue University, Montr\'eal, QC H3A 2A7, Canada}
\email{sethrsiegel@gmail.com}

\author[orcid=0000-0001-6903-5074, sname=Sievers, gname=Jonathan]{Jonathan Sievers}
\affiliation{Perimeter Institute for Theoretical Physics, 31 Caroline Street N, Waterloo, ON N25 2YL, Canada}
\affiliation{Department of Physics, McGill University, 3600 rue University, Montr\'eal, QC H3A 2T8, Canada}
\affiliation{Trottier Space Institute, McGill University, 3550 rue University, Montr\'eal, QC H3A 2A7, Canada}
\email{jonathan.sievers@mcgill.ca}






\begin{abstract}
High redshift Fast Radio Bursts (FRBs) are expected to be extremely powerful probes of our Universe. However, while a significant number of FRBs are expected to exist at high redshift, detecting them has been difficult, with only a handful robustly confirmed at redshifts greater than one. In many other fields, gravitational lensing from galaxy clusters has enabled high redshift detections by magnifying background sources. In this work we forecast the populations of FRBs expected to be detected by CHIME and upcoming instrument CHORD, for blank fields and by lensing through a range of strong lensing galaxy clusters, based on existing, observationally driven cluster models. We find that the presence of a galaxy cluster of mass $M\geq5\times10^{14}\,M_\odot$ within the detection beam of a transit telescope will approximately double the rate of detected high redshift ($z\geq1$ CHIME, $z\geq2$ CHORD) FRBs for that beam. Consequently, we find that knowledge of cluster positions can be used by instruments like CHIME or CHORD in tandem with novel observational strategies to isolate a sample of high redshift FRBs with $\gtrsim50\%$ purity at rate of $\lesssim3$ per year. This would provide a statistically high redshift sample of mostly gravitationally lensed FRBs, that would be ideal candidates for optical follow-up, constraining the FRB--star formation relation and for use in cosmological studies including measuring $H_0$, characterising dark matter substructures and probing reionization.

\end{abstract}

\section{Introduction}\label{sec:intro}
Fast radio bursts (FRBs) hold tremendous potential as a probe of extragalactic astrophysics, due, in part to their transient nature which allows a line of sight to be observed without the source. Proposed applications range from measuring the Universe's expansion history, $H(t)$ (including $H_0$ \citep{li_strongly_2018, wucknitz_cosmology_2021, james_measurement_2022}), identifying dark matter \citep{munoz_lensing_2016, sammons_first_2020, leung_constraining_2022} and constraining the epoch of reionization \citep{caleb_constraining_2019,linder_detecting_2020, pagano_constraining_2021}. In each of these cases, high redshift FRBs are particularly valuable.

Typically FRB search strategies have no preferred direction, assuming instead that the bursts are relatively homogeneous on the sky. Galaxy clusters however, present a significant source of large scale inhomogeneity on the sky that could violate this assumption. As the largest gravitationally bound structures in our Universe, galaxy clusters are hosted by enormous dark matter halos ($\gtrsim10^{14}M_\odot$) which can cause significant, achromatic magnification through the action of gravitational lensing. Leveraging this magnification to observe distant sources has a long history in astronomy \citep[see][for a review]{kneib_cluster_2011}, dating back to the suggestions of \citet{zwicky_rotverschiebung_1933} nearly 100 years ago. Today, cluster lenses are employed across the electromagnetic spectrum to constrain the faint end of galaxy luminosity functions at high redshifts \citep[e.g.]{bradac_focusing_2009}, study the morphology of individual stars as distant as redshift six \citep{welch_highly_2022} and investigate cosmological conundrums like the 
$H_0$ tension and the dark energy equation of state \citep{collett_constraining_2012, wong_h0licow_2020}. As we will show galaxy clusters can also significantly increase the rate of observed FRBs, particularly high redshift FRBs, even in surveys with resolutions an order of magnitude larger than a clusters extent. Moreover we show that by targeting clusters, statistically high redshift samples of FRBs can built without appealing to host galaxy localisation which is a resource intensive method.

Determination of an FRB's redshift relies upon association of its sky localisation with a host galaxy \citep{bannister_detection_2017, chatterjee_direct_2017}. In the regime of high redshift this requires not only a highly precise interferometric localisation, but also deep optical imaging of the localisation region to identify the host \citep{ryder_luminous_2023, sharma_preferential_2024}. As a result of these strict requirements only a few FRBs have been confirmed at redshifts beyond one. The next generation of FRB observatories, such as the VLBI outrigger stations constructed for CHIME will greatly improve the precision of localisations \cite{lanman_chimefrb_2024, cassanelli_fast_2024, shah_repeating_2025, the_chimefrb_collaboration_catalog_2025}, however, there are strict S/N requirements that potentially make distant bursts difficult. 

Due to the low number of confirmed high redshift sources it is commonplace in FRB studies to try and isolate potentially high redshift FRBs in alternative ways, with the most obvious being DM. The established Macquart relation \citep{macquart_census_2020} between DM and redshift allows DM to serve as a proxy for distance and therefore high DM FRBs are often considered to be high redshift. Large DM variance from host and intervening structures however, severely limits our ability to accurately estimate redshifts from DM \citep{james_z--dm_2021, tendulkar_host_2017, ocker_large_2022}. Instead, we show here that selecting FRBs observed along sight lines passing through galaxy cluster gravitational lenses can be an effective way to gather a high redshift sample, as these lenses not only boost the overall rate of FRBs, but enhance the fraction of observed events which are from high redshifts.

In this work we forecast the distribution of FRBs that will be seen through galaxy clusters, and suggest strategies to optimise the number of high redshift FRBs in future observations. In \S \ref{sec:method} we detail how we combine the existing methods for FRB population modelling with known models for the lensing action and electron densities of clusters to yield the expected distribution of observed FRBs along these sightlines. In \S \ref{sec:results} we show the expected rates, z, DM and scattering distributions of FRBs observed with the Canadian Hydrogen Intensity Mapping Experiment (CHIME), and the upcoming Canadian Hydrogen Observatory and Radio Transient Detector (CHORD) for both lensed and unlensed (blank) fields in a single tied array beam. In \S \ref{sec:discussion} we discuss these results in the context of CHIME and CHORD and suggest novel strategies for building statistically high redshift samples. Finally, \S \ref{sec:conclusion} provides an overall summary and some concluding remarks.

\section{Method}\label{sec:method}             

To calculate the redshift and dispersion measures distributions we expect to observe, we expand upon the z-DM model developed by \cite{james_z--dm_2021}. A complete description of the original model can be found in the corresponding paper, however we will provide a brief summary here. The original model calculated the expected rate of observed, non-repeating FRBs for a given survey. This model is also equally valid for describing the distributions of the first observed burst from any FRB source, including repeaters provided they adhere to the same energy function as non-repeating bursts \citep{james_modelling_2023}. To also capture the behaviour of subsequent repetitions would require expanding the model as performed in \cite{james_modelling_2023}, as well as knowing the fraction of true repeaters amongst the FRB population which is currently uncertain. Therefore we restrict our considerations to the distribution of the first observed FRB from each progenitor, allowing our expectations to describe a down-selected distribution of both repeating and non-repeating bursts without knowing the fraction of repeating FRBs. The corresponding non-repeating formalism yields an expected rate given by
\begin{align}
    R =& \int dz \Phi(z)\frac{dV(z)}{d\Omega dz}\int d\text{DM}p(\text{DM}|z)\nonumber\\
    &\times\int dB \Omega(B)\int dw p(w)p(E>E_{\text{th}}(B,w,z,\text{DM})),\label{eq:totalRate}
\end{align}
where $\Phi(z)$ is the source evolution function, $dV(z)/d\Omega dz$ describes the comoving volume per steradian per redshift interval, $p(\text{DM}|z)$ is the extragalactic DM distribution, $\Omega(B)$ is the inverse beam shape, describing the solid angle contained by the beam at relative gain $B$, $p(w)$ is the distribution of FRB widths (or duration, including propagation effects) and $p(E>E_{\text{th}})$ is the FRB cumulative energy function. The energy threshold $E_\text{th}$, is calculated by transforming a telescope detection fluence threshold ($F_0$) as per \cite{marani_cosmological_1996}, but taking into account the effects of varying FRB width and beam attenuation as
\begin{align}
    E_{\text{th}} =& \left[\frac{1}{(1+z)^\alpha}\right]\frac{4\pi D_L^2(z)}{(1+z)^{2}}\Delta\nu \frac{F_0 \sqrt{\,w}}{B},\label{eq:threshold}
\end{align}
where $D_L(z)$ is the luminosity distance at redshift $z$, $\Delta\nu$ is the FRB bandwidth, $w$ is the width of the FRB at observation (in units of detector resolution time) and $B$ is the gain relative to the beam maximum. Within the square brackets is a k-correction term which is only present under the spectral index interpretation of $\alpha$ (i.e. $F\propto \nu^\alpha$) \citep{macquart_fast_2018, james_z--dm_2021}. Equations \ref{eq:totalRate} and \ref{eq:threshold} provide the general means to evaluate the observed rate of FRBs for some choice of $\Phi(z)$, $p(E)$, $p(w)$, $\Omega(B)$ and $p(\text{DM}|z)$ which we will discuss in more detail in \ref{subsec:clusterContributions}. As with other studies \citep[e.g.][]{shin_inferring_2022}, we assume the source evolution function of FRBs is some power law proportional to the cosmic star formation rate (CSFR) defined by \cite{madau_cosmic_2014}
\begin{align}
    \Phi(z) =& \left[(1+z)^\alpha\right]\frac{\Phi_0}{(1+z)}\left(\frac{\text{CSFR}(z)}{\text{CSFR}(0)}\right)^n\\
    \text{CSFR} =& 1.0025738 \frac{(1+z)^{2.7}}{1+\left(\frac{1+z}{2.9}\right)^{5.6}}\label{eq:CSFR}
\end{align}
where the square brackets denote, in this instance, the k-correction term for the alternative, rate interpretation, of the spectral index $\alpha$ (i.e. $\Phi\propto \nu^\alpha$). 

In this study we forecast results for both planned and in-development instruments. Typically these instruments will search for FRBs in a large number of synthesised beams (interferometric beams formed from the elements of a radio telescope array) which tile the telescope's primary beam shape. To simplify the application over a broad range of systems we calculate rates for a single synthesised beam in each survey, assuming a Gaussian beam shape, allowing us to deploy the greatly simplifying approximation of a constant area $\Omega(B)d\log B$ per increment $d\log B$ in log beam gain. Moreover the clusters of interest will likely be contained within the extent of a single synthesised beam and therefore can be considered to have a constant primary beam gain over their extent.

\subsection{Surveys}
We forecast the expected rates within tied-array beams formed on both blank and lensed fields for three surveys over the CHIME and CHORD instruments. In each case we consider the tied-array beams to be simple Gaussians in shape. In detail these synthesised beams will be more complicated, depending on the projection of the array's baselines into the U-V plane towards the source, i.e. they will have fine scale structure and be both frequency and direction dependent \citep{ng_chime_2017}. In practice however, we find that a basic Gaussian with a full width half maximum (FWHM) corresponding to $\sim\lambda/D$ (wavelength/effective aperture diameter) for the longest characteristic baseline in the array describes the observed CHIME population fairly well (this is similar to the beam shapes used in \cite{michilli_analysis_2021}). Therefore, given the complexity and computational overhead associated with the true beam models we continue to make use of the Gaussian approximation for all surveys. The beam FWHM, central frequency ($\nu_c$), bandwidth ($\Delta\nu$), and the fluence threshold ($F_0$) used for each survey in our modelling can be found in Table \ref{tab:surveyParams}

\subsubsection{CHIME}
The Canadian Hydrogen Intensity Mapping Experiment (CHIME) is a radio interferometer located in Penticton, British Columbia. CHIME comprises four, fixed, parabolic half cylinder reflectors, oriented in the N-S direction, each with 256 dual polarisation feeds which operate in the 400--800 MHz range \citep[see][for an overview]{the_chimefrb_collaboration_chime_2018}. The approximately $\sim 230$ square degree field off view is sparsely populated by 1024 tied array beams formed via FFT (Fast Fourier Transform) beamforming \citep{ng_chime_2017}, each with an average FWHM of $21.5'$, that are constantly searched for FRB signals. Empirical estimates of CHIME's completeness indicate that's its detection system is $90\%$ \citep{merryfield_injection_2023} complete above a 5 Jy\,ms fluence threshold. In addition to basic sensitivity CHIME/FRB also has a complex selection function \citep{merryfield_injection_2023} which is biased against high width events, be they smeared high--DM signals or scattered events. Given the cursory nature of the investigations in this paper we deliberately make no attempt to account for these second order effects.

\subsubsection{CHORD}
The Canadian Hydrogen Observatory and Radio Transient Detector is a next generation radio interferometer \citep[see][for an overview]{vanderlinde_canadian_2019}. The full CHORD array will be composed of 512 6m dishes in a rectangular grid, with dual polarisation feeds operating from 300--1500 MHz. CHORD-64, which is currently under construction alongside CHIME at Dominion Radio Astrophysical Observatory, will comprise 64 of the 512 dishes in the full array. Compared to CHIME's $\sim50$\,K system temperature the receiver temperature of CHORD (which is expected to dominate the system temperature) will be lower at only $30$\,K. Given the lower collecting area of CHORD-64, lower system temperature and broader bandwidth a nominal FRB completeness threshold can be extrapolated for CHORD-64 based on the estimate for CHIME for a similar search algorithm implementation. Assuming that the observed FRBs are broadband over the entire 1200\,MHz observed by CHORD-64, the CHIME threshold fluence transforms to $\sim2.5$\,Jy\,ms for CHORD-64. Assuming the dishes are placed in a regular square grid the average FWHM of a CHORD-64 tied-array beam will be ($\sim\lambda/D = $ 0.33\,m / ($6$\,m $\times$ 8)$\approx 23.9'$). Conversely the full CHORD-512 array will have a collecting area and sensitivity $\times8$ better at $F_0\approx 0.3125$\,Jy\,ms, and a tied-array beam FWHM $\times\sqrt(8)$ smaller at $8.45'$.

\begin{table}
    \setlength{\arrayrulewidth}{0.5mm}
    \centering
    \renewcommand{\arraystretch}{1.2} 
    \setlength{\tabcolsep}{8pt} 
    \begin{tabular}{|c||>{\centering\arraybackslash}p{2.5cm}|>{\centering\arraybackslash}p{2.5cm}|>{\centering\arraybackslash}p{2.5cm}|>{\centering\arraybackslash}p{2.5cm}|}
        \hline
        \textbf{Surveys} & Beam FWHM (arcmin) & $\nu_C$ (MHz) & $\Delta\nu$ (MHz) & $F_0$ (Jy\,ms) \\ \hline \hline
        CHIME & 21.5 & 600 & 400 & 5 \\ \hline
        CHORD-64 & 23.9 & 900 & 1200 & 2.5 \\ \hline
        CHORD-512 & 8.45 & 900 & 1200 & 0.3125 \\ \hline
    \end{tabular}
    \caption{Survey Parameters.}
    \label{tab:surveyParams}
\end{table}

\subsection{Cluster Contributions}\label{subsec:clusterContributions}

The form of the remaining functions, $p(E)$, $p(\text{DM}|z)$ and $p(w)$ will depend on the lensing, scattering and dispersion contributed by the morphology of any foreground cluster. To account for this morphology dependence in our treatment of cluster lensing we calculate $p(E)$, $p(w)$ and $p(\text{DM}|z)$ as a function of redshift and sky position (or beam gain in the formalism of Eq. \ref{eq:totalRate}) for ten clusters of varying mass, using existing models for their lens potentials and their electron density profiles.

\subsubsection{Cumulative Energy Function}
Intrinsically $p(E>E_{th})$ is often assumed to have a power-law or Gamma function form (Schecter function), for simplicity we make use of a power law model with a hard cutoff. In a lensing context this only introduces marginal differences from the exponential roll off of the Gamma function \citep{sammons_effect_2022}.

Magnification from gravitational lensing by a cluster increases the observed fluence from an FRB, allowing bursts of energy $E > E_{\text{th}}/\mu$ to be observed, where $\mu$ is the magnification factor. As a result the lensed FRB cumulative energy function becomes 
\begin{equation}
    p_L(E>E_{\text{th}}) = \int d\mu\,p_S(\mu)p\left(E>\frac{E_{\text{th}}}{\mu}\right),
\end{equation}
which may be expressed as the log space convolution of the intrinsic FRB cumulative energy function, with $\mu p_S(\mu)$, where $p_S(\mu)$ describes the probability of observing a source with magnification $\mu$. 

To determine the probability of source magnification through a given cluster we must have a model for the cluster's gravitational potential. The gravitational potential of the cluster causes deflection of incident radiation which ultimately results in the gravitational lensing we observe. As the path length through the cluster is often negligible compared to the distance to the source, the lensing potential can be approximated by projecting the cluster's Newtonian potential ($\Phi$) onto the image plane

\begin{equation}\label{eq:lensPotential}
    \psi(\Vec{\theta}) = \frac{2}{c^2}\frac{D_{ds}(z_d,z_s)}{D_s(z_s)D_d(z_d)}\int\Phi(D_d\Vec{\theta},\ell)d\ell,
\end{equation}
where $D_s$, $D_d$, $D_{ds}$ are the angular diameter distances from the observer to the source at redshift $z_s$, the lens at redshift $z_d$, and from the lens to the source respectively, $\Vec{\theta}$ is a 2D vector describing a point on the image plane and $\ell$ is the path through the cluster. 

A ray incident upon the lens potential at point $\Vec{\theta}$ is then deflected by angle $\Vec\alpha = \nabla\psi(\Vec{\theta})$ and an observer will see images of a point in the source plane $\Vec{\beta}$ for all $\Vec{\theta}$ satisfying the lens equation 
\begin{equation}\label{eq:lensEquations}
    \Vec{\alpha}=\nabla\psi(\Vec{\theta}) = \Vec{\theta}-\Vec{\beta}.
\end{equation}

The magnification of each image is given by
\begin{equation}\label{eq:magni}
    \mu = \frac{1}{|(1-\kappa)^2 - \gamma^2|},
\end{equation}
where $\kappa$ and $\gamma$ are known as the convergence and shear, and defined for each point in the image plane ($\theta_{ij}$) as 
\begin{equation}\label{eq:kappagamma}
    \kappa\equiv\frac{1}{2}\nabla^2\psi,\qquad \gamma\equiv \sqrt{\left(\frac{1}{2}\frac{\partial^2\psi}{\partial\theta_i^2}-\frac{1}{2}\frac{\partial^2\psi} {\partial\theta_j^2}\right)+\left(\frac{\partial^2\psi}{\partial\theta_i\partial\theta_j}\right)^2}.
\end{equation}

The probability of observing a source at specific redshift with a magnification $\mu$ can then be determined using the values of $\kappa$ and $\gamma$ to evaluate the magnification at each point across the image plane. Each magnification must map to an area proportional to $1/\mu$ in the source plane and therefore for sources with a statistically uniform distribution in the source plane, the probability of magnification $\mu$ can be approximated as proportional to the number of points with magnification $\mu$ in the image plane, weighted by $1/\mu$. This method is only an approximation of the true probability distribution as it assumes each source has only one image, which is not necessarily true. However, calculating the true distribution is computationally intensive and we find that it yields only marginally different results when compared to the approximation detailed above, therefore we will only use the single image approximation of $p_S(\mu)$ moving forward. 

Many surveys have been conducted to constrain the lens models of galaxy clusters. Notable among them include the sample from the Hubble Frontier Fields (HFF) \citep{johnson_lens_2014}, the Cluster Lensing and Supernova Survey with Hubble (CLASH) \citep{postman_cluster_2012} and the Reionization and Lensing Cluster Survey (RELICS) \citep{coe_relics_2019}, which have each made the relevant lensing models available online through the Minkulski Archive for Space Telescopes (MAST). A range of methods have been developed to derive these lensing models, with significant discrepancies between them. As such multiple lensing models corresponding to the different methods are usually available for any given cluster. For consistency we only make use of models derived using \texttt{LENSTOOL} \citep{kneib_lenstool_2011} as this method employs a multi-scale approach which allows the model to be defined over the relatively large angular extent we require\footnote{Typically the size of the synthesised telescope beam we are considering is substantially larger than the angular extent of a cluster.} without sacrificing small scale structure. We also quantify the impact of lens model uncertainty by calculating FRB rates for four different models of the same cluster derived using different methods in appendix \ref{app:magUncertainty}. We find that  differences in the morphology of the magnification distribution resulting from the choice of method result in second order differences in our results, which do not impact the conclusions we draw in this work.

\begin{figure}
    \centering
    \includegraphics[width=0.8\linewidth]{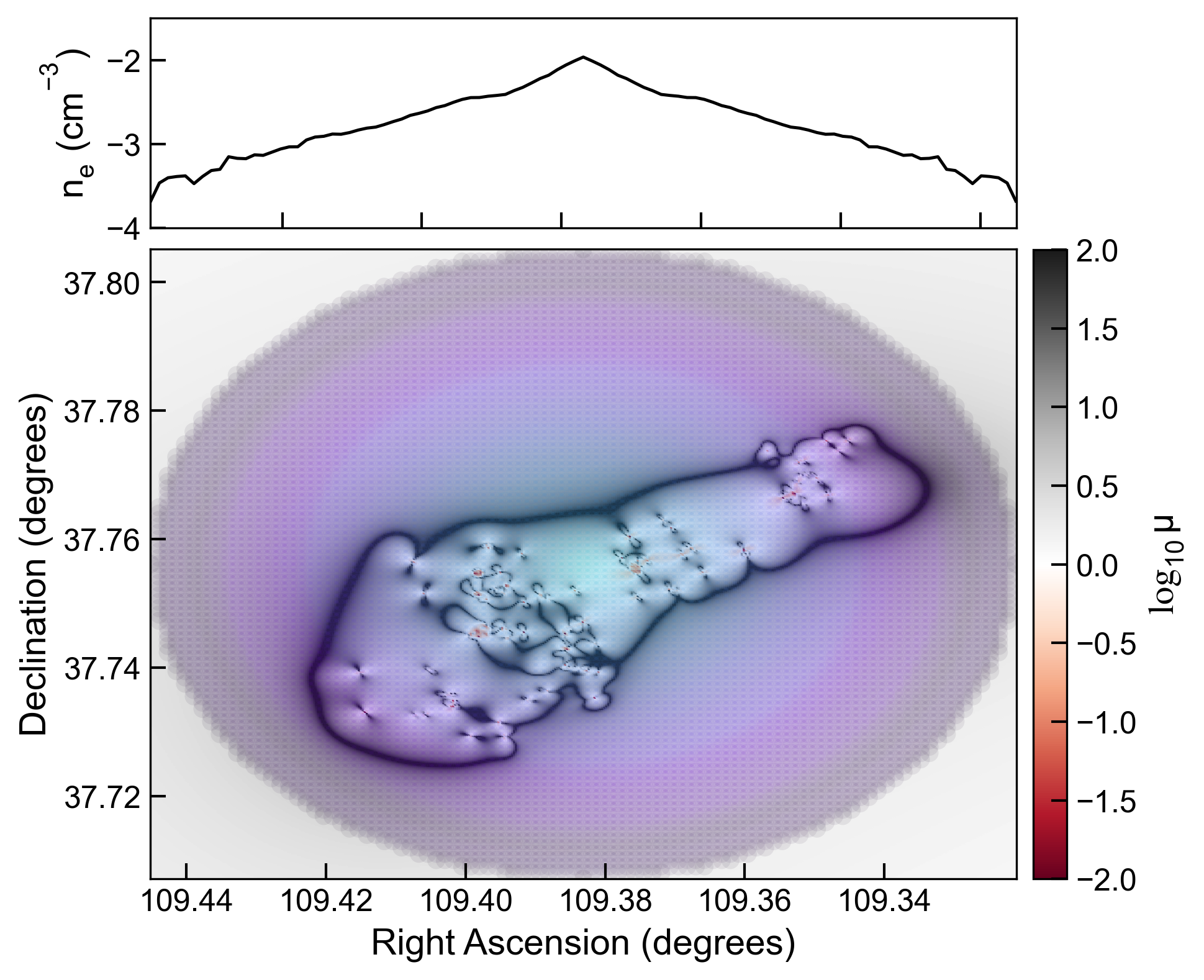}
    \caption{Cluster morphology overlap plot for MACS J0717.5+3745. Black and red heatmap displays on a log scale the best fit gravitational lens magnification model \citep{johnson_lens_2014} overlaid on the violet heatmap of the 2D projection of the radial electron density map (top) derived from x-ray observations \citep{cavagnolo_intracluster_2009}.}
    \label{fig:NeMuOverlap}
\end{figure}

As a part of each cluster lens model $\kappa$ and $\gamma$ are provided. As seen from equations \ref{eq:lensPotential} and \ref{eq:kappagamma} these values depend on the source redshift, varying linearly with the distance ratio $D_{ds}(z_d,z_s)/D_s(z_s)$. The archived models provide these quantities with $D_{ds}(z_d,z_s)/D_s(z_s)=1$ such that they can be scaled simply to any source redshift. We then combine the scaled $\kappa$ and $\gamma$ following Eq. \ref{eq:magni}, to yield the source magnification as a function of sky position for sources at redshift $z_s$. An example of such a map can be seen in Fig. \ref{fig:NeMuOverlap} overlaid on a map of ICM electron density for MACS J0717.5+3745, one of the largest clusters known. The magnifications are calculated for the default case $D_{ds}(z_d,z_s)/D_s(z_s)=1$, which typically yields magnified regions of the largest extent. Each source magnification map is defined on a regular grid consistent with the input $\kappa$ and $\gamma$ maps. Given the large variance in the high magnification limit between models, we follow \cite{johnson_lens_2014} and limit the maximum magnification to 100. At radio frequencies wave optics can also play a role in restricting the maximum magnification. Typically considerations of wave effects have been in the context of multiple images \citep[see]{leung_wave_2023} however the maximum magnification of a single image will also be affected if the size of that image exceeds the Fresnel scale in the image plane. The Fresnel scale is $r_F=\sqrt{D_{\text{eff}\lambda}}$, where $D_\text{eff}$ is the effective angular diameter distance $D_dD_s/D_{ds}$ and $\lambda$ is the wavelength in the frame of the lens. For a cluster at redshift 0.5 and a source at redshift one $r_F\sim10^9\,$km. At our maximum magnification of $\mu=100$ any emission region smaller than $10^7\,$km will not have an image large enough to be meaningfully influenced by wave optics. Amongst the currently favoured magnetar progenitor models, shock type emission yields the largest emission regions \citep{margalit_constraints_2020} which are still expected to be smaller than $10^7\,$km \citep{nimmo_magnetospheric_2024}, therefore we do not consider wave optics further.

The rate of FRB detections is impacted by dispersion and scattering along the line of sight as seen in Eq. \ref{eq:totalRate}. Therefore in order to accurately calculate the impact of lensing on detected rates it is important to model the overlap between $n_e$ and $\mu$ maps. From Fig. \ref{fig:NeMuOverlap} it is apparent that a cluster's $n_e$ and $\mu$ distributions can be morphologically unique justifying the complex, cluster specific modelling approach we use here.

We construct the probability of source magnification, $p_S(\mu, B, z_s)$, at each redshift from the normalised weighted histogram of magnifications within each constant area annulus of $\Omega(B)$ with gain $B\pm d\log B$. Each magnification is weighted by $1/\mu$ to account for the reduced area in the source plane corresponding to the magnified region, yielding $p_S\propto\mu^{-3}$ in the high magnification limit as expected \citep{rauch_gravitational_1991}. Regions on the image plane within the beam model but outside the lens model are assumed to have magnification $\mu=1$. Given the degeneracy in detected fluence between magnification and beam gain it is crucial that the uncertainty in beam attenuation applied to a given location in Fig. \ref{fig:NeMuOverlap} is significantly lower than the magnification at that point \footnote{We highlight that this is a condition we impose for a self consistent model and is separate from considerations of the true synthesised beam shape.}. Therefore we opt to calculate $p_S(\mu, B, z_S)$ distributions for 100 beam annuli between $B=1$ (zenith) and $B=10^{-3}$, yielding a maximum gain error of $7\%$ for the simple Gaussian beam model, small compared to the range of potential magnifications.

\subsubsection{DM Distribution}
An FRBs extragalactic\footnote{As in previous studies we neglect the foreground Milky Way contribution to an FRBs observed DM as it is relatively small and well constrained, allowing for more reliable subtraction.} DM typically comprises two components
\begin{equation}
    \text{DM}_{\text{EG}} = \text{DM}_{\text{Cosmic}}+\frac{\text{DM}_{\text{Host}}}{(1+z)},
\end{equation}
where DM$_{\text{Cosmic}}$ is the contribution from the diffuse intergalactic medium (IGM) and foreground galaxy halos, and DM$_{\text{Host}}$ is the contribution of the host galaxy, scaled from its rest frame at redshift $z$ to the observers frame. The probability distribution of DM$_\text{EG}$ is then the convolution of the independent probability distributions for $\text{DM}_{\text{Cosmic}}$ and DM$_{\text{Host}}$. The distribution of $\text{DM}_{\text{Cosmic}}$ follows the formalism of the Macquart relation \citep[see][]{macquart_census_2020, james_z--dm_2021}, with the mean DM$_{\text{Cosmic}}$ value being 
\begin{equation}\label{eq:DMCosmic}
    \langle \text{DM}_{\text{Cosmic}}\rangle = \int\limits_0^z\frac{c\bar{n}_edz'}{H_0(1+z')^2E(z)},
\end{equation}
where $\bar{n}_e\approx 0.2\,$m$^{-3}$ is the mean cosmological electron density at redshift zero. Whereas the host contributions are described by an empirical log-normal distribution with a mean $\mu^*_{\text{Host}}=2.27$ and standard deviation $\sigma^*_{\text{Host}}=0.55$, as fit in \cite{james_measurement_2022}.

Clusters are also expected to host a significant reservoir of free electrons in their intra-cluster mediums (ICM)\footnote{We highlight that this is distinct from the diffuse IGM which contributes to DM$_\text{Cosmic}$ in Eq. \ref{eq:DMCosmic}}. The electron density profiles of clusters can be inferred from x-ray observations of the ICM. We make use of the radial electron density profiles available on the Archive of \textit{Chandra} Cluster Entropy Profile Tables (ACCEPT), derived via uniform analysis of archival \textit{Chandra} X-ray data \citep{cavagnolo_intracluster_2009}. From the radial profiles we re-construct a circularly symmetric 2D electron density map centred at the clusters position as seen in Fig. \ref{fig:NeMuOverlap}. As expected the density of electrons is substantial, ranging from IGM-like ($10^{-4}$ cm$^{-3}$) at the edges to ISM-like ($10^{-2}$ cm$^{-3}$) in their cores. Furthermore the electron density profile is much smoother than the corresponding magnification map, as the hot ICM, which dominates the free electron content, cannot form the same the small scale features which are prevalent in the overall mass distribution. Notably, Fig. \ref{fig:NeMuOverlap} shows that the majority of excess magnification will come from regions of high electron density, highlighting the importance of accounting for the contributions of the cluster DM and scattering in our calculations. 

As a result of the large electron density and characteristic scale of galaxy clusters ($\sim1$ Mpc), any FRB piercing them is expected to pick up a large DM$_{\text{Cluster}}$ contribution. For each position in Fig. \ref{fig:NeMuOverlap} we calculate an associated DM$_{\text{Cluster}}$ and scattering contribution, assuming a uniform cluster depth of 1\,Mpc in all cases. In reality the cluster depth will of course depend on the 3D structure of the cluster. For MACS J0717.5+3745 we recalculate the expected lensed z--DM distribution using a projected 3D model of the electron distribution based on observations of the Sunyeav Zeldovich effect and compare with the 1\,Mpc depth, x-ray profile based approached described above and find less than $10\%$ difference in the detected rate as a function of redshift, but up to a factor of two difference in the expected DM distributions (see appendix \ref{app:3DEDModel}). Given current uncertainties in population parameters and other DM contributions however, a more realistic model of cluster depth is not expected to improve the precision of our results, hence we opt for simplicity. We highlight however that for applications such as probing reionization, for which the DM distribution is crucial, cluster electron profiles will require more precise modelling \citep[see]{breuer_thermodynamic_2025}. Given a probability distribution of DM$_{\text{Cluster}}$ this cluster contribution can be folded into the existing formalism just as for DM$_{\text{Host}}$ by convolving the probability of DM$_\text{Cluster}$ at a given beam gain and redshift, $p(\text{DM}_{\text{Cluster}}|B,z_S)$, with the $p(\text{DM}_\text{EG})$, formed from the cosmic and host probability distributions, at each redshift and beam gain increment. 

Similarly to the probability of source magnification, we construct $p(\text{DM}_{\text{Cluster}}|B,z_S)$ by normalising the histogram of positions from redshift $z_S$ detected within concentric annuli of the beam of gain $B\pm d\log B$. However, where in the previous case of source magnification, area in the source plane was weighted uniformly, in the case of DM, FRB number counts are weighted uniformly. At a position with magnification $\mu$ there are $1/\mu$ fewer FRB sources at redshift $z_S$, but $1/\mu^{\gamma}$ more detectable FRB sources owing to the energy function of FRBs providing more low energy bursts, resulting in weights at each position given by $1/\mu^{1+\gamma}$.

\subsubsection{Width Distribution}
The distribution of FRB widths, as observed by the incoherent detection algorithm assumed in Eq. \ref{eq:threshold}, depend not only on the intrinsic distribution of widths set by the unknown FRB emission mechanism, but also on propagation effects such as multipath scattering and instrumental effects such as DM smearing. In treating observed widths, the z--DM formalism follows the effective width model from \cite{cordes_searches_2003}
\begin{equation}
    w_{\text{eff}} = \sqrt{w_{\text{int}}^2+w_{\text{scatt}}^2+w_{\text{DM}}^2+w_{\text{samp}}^2+w_{\text{Cluster}}^2},
\end{equation}
where $w_{\text{int}}$ is the intrinsic width, $w_{\text{scatt}}$ is the additional width from multipath scattering, $w_{\text{DM}}$ is the width contributed by DM smearing and $w_{\text{samp}}$ is the sampling time. In general the interaction between different width contributions is non-trivial and will not be captured by this model. Instead this model aims simply to define an effective width which will be dominated by any large contribution. In keeping with this model we include the scattering time contributed by a cluster ($w_{\text{Cluster}}$) as an additional term in the above equation and determine the probability distribution of $w_\text{eff}$ empirically from monte-carlo simulations of the underlying distributions. As in \cite{james_z--dm_2021} $w_\text{DM}$ is set by DM and channel width, $w_\text{samp}$ is considered to be constant and together they set the minimum width at a given frequency. Both $w_\text{int}$ and $w_\text{scatt}$ are considered to be distributed as log-normals with parameters described in Table \ref{tab:inputs} and the probability of a given cluster scattering time $p_{\text{Cluster}}(\tau|B,z_S)$, is derived from the specific cluster morphology. We calculate $p_{\text{Cluster}}(\tau)$ identically to $p(\text{DM}_{\text{Cluster}}|B,z_S)$, except rather than DM, the scattering at each foreground position is calculated as \citep[from][for a diffractive scale less than the inner scale of the turbulence]{macquart_temporal_2013} 
\begin{align}
    w_{\text{Cluster}} &= 4.1\times10^{-5}\frac{1}{(1+z_d)}\left(\frac{\lambda_0}{1\text{m}}\right)^4\left(\frac{\ell_0}{1\text{AU}}\right)^{1/3}\\
    &\times\left(\frac{D_\text{eff}}{1\text{Gpc}}\right)\left(\frac{\text{SM}_{\text{eff}}}{10^{12}\text{m}^{-17/3}}\right)\text{s},\nonumber\\
    \text{SM}_{\text{eff}} &= 8.4\times10^{-13}\left(\frac{n_e}{10^{-4}\text{cm}^3}\right)^2\left(\frac{L_0}{0.1\,\text{Mpc}}\right)^{-2/3}\\
    &\times\frac{\Delta L}{(1+z_d)^2}\,\text{m}^{-17/3},\nonumber
\end{align}
where $z_d$ is the redshift at the deflection (cluster), $\lambda_0$ is the wavelength in the observers frame, $L_0$ and $\ell_0$ are the outer and inner turbulence scales, SM$_{\text{eff}}$ is the effective scattering measure, where 
the dimensions of 17/3 derive from the amplitude of turbulence per unit length integrated along the line of site for a density power spectrum following a kolmogorov turbulent cascade, and $\Delta L$ is the cluster's depth. Similarly to assuming a fixed cluster depth $\Delta L$=1\,Mpc, we also assume fixed turbulence scales, with $L_0$ left at the default value of $0.1\,$Mpc and $\ell_0$ being set by the scale of viscous dissipation, which we take to be fixed at $\sim10\,$kpc based on magneto-hydrodynamical simulations of Coma-like clusters \citep{gaspari_constraining_2013}. We note that while this value has a large spread amongst ICM turbulence literature \citep[1--100$\,$kpc values seen in][]{zhang_mapping_2024, cucchetti_towards_2019}, the shallow dependence of scattering time on the inner scale of turbulence somewhat mitigates the impact of this uncertainty.

\begin{table*}
    \setlength{\arrayrulewidth}{0.5mm}
    \centering
    \begin{tabular}{|c|c|c|}
        \hline
        \textbf{Parameter} & \textbf{Input Value} & \textbf{Explanation}\\
        \hline
        $E_{\text{max}}$  & $10^{41.7}$ erg & Maximum FRB energy\\
        $E_{\text{min}}$ & $10^{30}$ erg & Minimum FRB energy\\
        $n_\text{sfr}$  & 1.15 & Cosmic star formation rate scaling index \\
        $\alpha$ & -1.03 & Spectral Rate index \\
        $\gamma$  & -0.95 & Energy function index \\
        $\mu_\text{Host}$ & 2.23 & Mean of the normal $\log_{10}$DM$_{\text{Host}}$ distribution in pc cm$^{-3}$\\
        $\sigma_\text{Host}$ & 0.57 & Standard deviation of the normal $\log_{10}$ DM$_{\text{Host}}$ distribution in pc cm$^{-3}$\\
        $\Phi_0$ & $2.43\times10^4$ Gpc$^{-3}$yr$^{-1}$& Volumetric FRB rate at $z=0$ (see caption) \\
        \hline
    \end{tabular}
    \caption{Input parameters for z--DM population, all parameters from \citet{james_measurement_2022} from best fits assuming flat priors on all parameters, with the exception of $\Phi_0$ which we set to approximately yield the observed CHIME rate as described in the text and $E_{\text{max}}$ which was increased by \citet{ryder_luminous_2023}. Other parameters not listed here retain their default values which can be found in the public z--DM implementation and in \citet{james_measurement_2022}.}
    \label{tab:inputs}
\end{table*}

\section{Results}\label{sec:results}
\begin{figure*}
    \includegraphics[width=\textwidth]{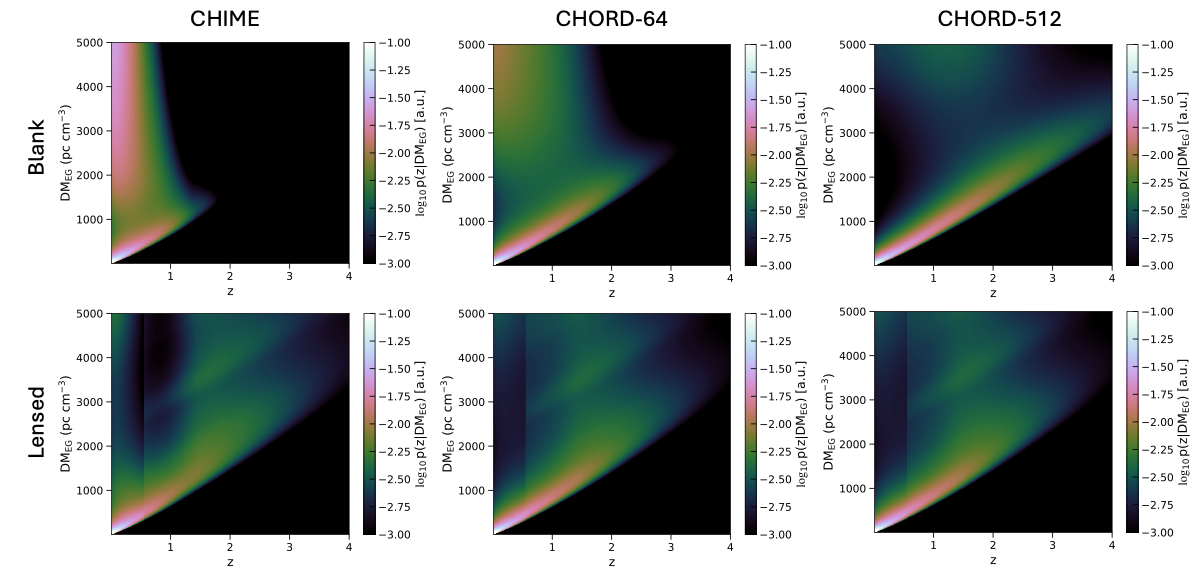}
    \caption{The z-DM distribution expected to be observed in a single tied array beam by the CHIME/FRB (\textit{left}), CHORD-64 (\textit{middle}) and CHORD-512 (\textit{right}) instruments based on the survey specifications and best fit population parameters detailed in Tables \ref{tab:surveyParams} and \ref{tab:inputs} respectively. \textit{Top} shows the z--DM distributions expected in the case of a blank field. \textit{Bottom} shows the z--DM distributions expected when a synthesised beam is centred on cluster MACS J0717.5+3745 at $z=0.545$
    described in Table \ref{tab:clusters} and depicted in Appendix \ref{app:clusters}. The sharp feature seen in the lensed plots is onset of lensing at the cluster redshift and represents a physical step change at the $\Delta z=0.01$ resolution used here}
    \label{fig:combined}
\end{figure*}

To assess the impact of a cluster on the observed FRB population we first evaluate the blank field z--DM distribution and absolute rate expected for FRBs observed by CHIME and CHORD in both its 64 dish and full 512 dish deployment configurations, as seen in the top row of Fig. \ref{fig:combined}. Shown in the Figure is the conditional probability of detecting an FRB from redshift $z$ given it has a dispersion DM$_{\text{EG}}$, $p(z|\text{DM}_{\text{EG}})$. We highlight that previous applications of z--DM often plot the two-dimensional probability $p(z,\text{DM}_{\text{EG}})$ density, where instead we show $p(z,\text{DM}_{\text{EG}})/p(\text{DM}_{\text{EG}})$. Each of these plots show a mixture of common trends which are well understood and more unique features which are the focus of this study. Common to each is the minimum DM$_{\text{EG}}$ at each redshift associated with DM contributed by diffuse gas in the IGM, i.e. the Macquart relation \citep{macquart_census_2020}. Secondarily each plot also shows large maximum DMs at each redshift associated with excess DM contributions from the host galaxy or other intervening structures such as a cluster, the precise effect of which we discuss in the forthcoming sections.

Unlike previous applications of z--DM to ASKAP, FAST and DSA, which constrain the population parameters using observations from the relevant instruments, we highlight that these estimates are strictly forecasts, allowing uniform comparison between instruments, using the best fit population parameters from previous studies and information about those instruments. Nonetheless, these estimates are the first of their kind for CHORD, providing useful forecasts even in the unlensed case, and the first for CHIME using the z--DM code \citep[previous treatments like \cite{shin_inferring_2022} use a different implementation of the same method]{prochaska_frbszdm_2023}, providing a useful point of comparison.

A direct comparison of our CHIME expectations with previous measurements by \cite{shin_inferring_2022} is difficult as they use a more complex treatment of CHIME sensitivity informed by injections. Despite this difference we see good agreement, with both distributions yielding approximately the same observable range for CHIME in z--DM space\footnote{The cardinal difference is a small excess probability at higher redshifts in the \cite{shin_inferring_2022} model due to the use of a Gamma function energy distribution as opposed to the power law we apply here.}. This agreement between the z--DM distribution expected from CHIME, based on the best fit luminosity function found by ASKAP, and the z--DM distribution fit to CHIME observations indicates that our chosen prescription captures the behaviour of FRB observations sufficiently well for our study on the relative effect of lensing. To yield an expected rate we fit the total rate of FRBs per Gpc$^{-3}$ per year ($\Phi_0$) to match the 474 apparently non-repeating FRBs observed over the 214.8 day duration of the first CHIME FRB survey \citep{the_chimefrb_collaboration_first_2021}. This rate of $\approx 2.43\times10^4\,$Gpc$^{-3}$yr$^{-1}$, is consistent within 1.3$\sigma$ with the results of \cite{shin_inferring_2022}. Moreover this roughly derived rate will be an underestimate as we do not account for primary beam attenuation or non-linear changes in the rate over CHIME's large dynamic range in bandwidth, which due to a negative spectral index will also increase the rate. These are expected to be order unity discrepancies, and are therefore ignored for simplicity as we are primarily interested in relative change in observed rates due to lensing, however they will also make our resulting rates for each instrument conservative.

Extending to CHORD, Fig. \ref{fig:combined} shows that CHORD-64 will observe deeper than CHIME, as expected given its lower detection threshold. Furthermore, the full CHORD-512 will observe again more deeply, potentially detecting FRBs beyond a redshift of 4. Notably, for CHORD-512 these high redshift events are expected to have similar DM's to the highest DM candidates found at lower redshift, with the $p(z|\text{DM}_{\text{EG}})$ flattening to become more uniform at higher DMs. In general this means that a larger fraction of high DM events observed by CHORD-512 will be genuinely higher redshift compared to CHORD-64 and CHIME, the high DM samples of which will be dominated by events with high host DM contributions. However, even for CHORD-512 it remains the case that a sample of high redshift FRBs will be difficult to identify based on DM alone, presenting an opportunity for cluster lines of sight to provide an alternative way to isolate a sample of high redshift FRBs. 

From amongst the available, well modelled clusters contained in Hubble Frontier Fields (HFF), Cluster Lensing And Supernova survey with Hubble (CLASH) and Reionization Lensing Cluster Survey (RELICS) samples we take five of the six clusters in the HFF sample that have electron density profiles available on the CHANDRA Archive, as these clusters are some of the most well studied cases of cluster lensing due to their high masses. To capture a wide range of lensing behaviour the remaining five clusters in our sample are comprised of clusters with the lowest Sunyeav-Zeldovich masses (as measured by the Planck experiment in their 2015 data release \citep{ade_planck_2016}) for which there are also lens models and electron density profiles available. A majority of the candidates satisfying these criteria come from the RELICS survey, and hence for consistency of lensing models we neglect any candidates from CLASH in our sample. The properties of the clusters included in our sample are detailed in Table \ref{tab:clusters} and the corresponding lens models can be obtained from the MAST archive at \texttt{https://archive.stsci.edu/prepds/frontier/} or \dataset[doi:10.17909/T9SP45]{https://dx.doi.org/10.17909/T9SP45}. The masses across the sample vary between $\approx 5\times10^{14}\,M_\odot$ and $\approx 11\times10^{14}\,M_\odot$ corresponding to the least and most massive clusters in the CLASH, RELICS and HFF surveys. The sample redshifts also show some variation across the range expected for massive clusters ($\sim 0.2$--$0.5$) \citep{chiu_erosita_2022}. The morphologies of the lensing and electron density models for the selected clusters are depicted in Appendix \ref{app:clusters} and display a variety of lensing morphologies and radial electron density profiles in both the high and low mass regimes. While this small sample does not exhaustively sample the mass--redshift--morphology parameter space of galaxy clusters, it captures a sufficient range of cluster properties to allow us to discern whether cluster lensing will have a significant impact on observed FRB populations, or whether only the most extreme clusters will introduce substantial lensing effects.

\begin{table}
    \setlength{\arrayrulewidth}{0.5mm} 
    \centering
    \renewcommand{\arraystretch}{1.2} 
    \setlength{\tabcolsep}{8pt} 
    \begin{tabular}{|c||c|c|>{\centering\arraybackslash}p{2.5cm}|>{\centering\arraybackslash}p{2.5cm}|>{\centering\arraybackslash}p{2.5cm}|}
        \hline
        \textbf{Cluster Name} & \textbf{Mass ($10^{14}\,$M$_\odot$)} & \textbf{Redshift} & \multicolumn{3}{c|}{\textbf{Rate per 1000 tied-array beam days (rate $z\geq1$)}$^*$} \\ \hline  
        & & & CHIME & CHORD-64 & CHORD-512 \\ \hline \hline
        Blank & -- & -- & 2.19 (0.10) & 7.73 (1.58) & 14.13 (6.13) \\ \hline
        MS 1008.1-1224 & $4.94^{+0.57}_{-0.60}$ & 0.306 & 2.27 (0.15) & 7.93 (1.73) & 16.27 (7.73) \\ \hline
        Abell 2537 & $5.52^{+0.51}_{-0.51}$ & 0.297 & 2.43 (0.28) & 8.27 (1.98) & 21.60 (12.07) \\ \hline
        MACS J0257.1-2325 & $6.22^{+0.70}_{-0.74}$ & 0.505 & 2.25 (0.14) & 7.93 (1.70) & 15.67 (7.40) \\ \hline
        MACS J0035.4-2015 & $7.01^{+0.45}_{-0.50}$ & 0.352 & 2.34 (0.21) & 8.11 (1.85) & 18.20 (9.40) \\ \hline
        RXC J0232.2-4420 & $7.54^{+0.33}_{-0.32}$ & 0.284 & 2.40 (0.25) & 8.20 (1.92) & 19.13 (10.00) \\ \hline
        Abell 370 & $7.64^{+0.56}_{-0.57}$ & 0.375 & 2.53 (0.36) & 8.60 (2.21) & 24.00 (14.27) \\ \hline
        Abell 2744 & $9.84^{+0.38}_{-0.39}$ & 0.308 & 2.76 (0.47) & 9.40 (2.77) & 25.60 (14.80) \\ \hline
        MACS J1149.5+2223 & $10.41^{+0.52}_{-0.55}$ & 0.543 & 2.30 (0.20) & 8.07 (1.87) & 17.73 (9.40) \\ \hline
        AS1063 & $11.36^{+0.34}_{-0.34}$ & 0.348 & 2.37 (0.23) & 8.27 (1.96) & 18.27 (9.40) \\ \hline
        MACS J0717.5+3745 & $11.48^{+0.53}_{-0.55}$ & 0.545 & 2.73 (0.57) & 9.20 (2.83) & 26.73 (17.60) \\ \hline
    \end{tabular}
    \caption{\vspace{5pt} Single tied-array beam rates for CHIME, CHORD-64, and CHORD-512 in blank and lensed fields.
    \newline
    $^*$We caution that the lensed rates may not be observed every day by simply forming 1000 beams simultaneously, as not all formed beams may contain clusters.}
    \label{tab:clusters}
\end{table}

For each of the clusters we calculate the expected z--DM distribution and total rate of FRBs observed within a single synthesised beam of the CHIME, CHORD-64 and CHORD-512 telescopes centred on the cluster's location. The resulting rates are listed in Table \ref{tab:clusters}. For comparison and to highlight the general effects of cluster lensing on the observed z--DM distributions, we include the most extreme case of lensing in our sample alongside the blank field cases in the bottom and top rows of Fig. \ref{fig:combined} respectively. The remaining z--DM distributions and corresponding morphologies are depicted in appendix \ref{app:clusters}. This case corresponds to the lensing expected from MACS J0717.5+3745 ($z=0.545$; hereafter MACS 0717), one of the most massive clusters ever observed as well as one of the largest gravitational lenses \citep{johnson_lens_2014, zitrin_largest_2009}. For each telescope the lensed z--DM distributions depicted in Fig. \ref{fig:combined} show a sharp increase in the $p(z|\text{DM}_{\text{EG}})$ at redshifts greater\footnote{We highlight that while the conditional probabilities are shifted in the region below the cluster's redshift, due to the change in overall behaviour, the absolute properties of the z--DM distribution below the cluster redshift are identical to the case of the blank field.} than 0.545, the redshift of MACS 0717. This increase in $p(z|\text{DM}_{\text{EG}})$, caused by magnification from gravitational lensing by the cluster, extends the observed z--DM distribution to higher redshifts and DMs than in the case of blank fields. In the extreme case of MACS 0717 the redshift range accessible by each instrument is greatly boosted. Marginalising the two-dimensional probability density over the DM axis, the lensing from MACS 0717 increases the $[0.5,0.95,0.99]$ confidence intervals from redshifts $[0.37, 1.15, 1.52]$ to $[0.48, 1.96, 2.96]$ for CHIME, from $[0.34, 1.12, 1.6]$ to $[0.45, 1.97, 2.99]$ for CHORD-64 and from $[0.77, 2.23, 3.11]$ to $[1.29, 2.92, 3.93]$ for CHORD-512. The maximum DM at redshifts above the cluster is also greatly increased due to contributions from the cluster itself, which are of $\mathcal{O}(10^3)$ pc cm$^{-3}$. Conversely, the lensed distribution retains the same maximum redshift at a given DM, as this is specified by the diffuse IGM component along all lines of sight independently of cluster presence. 

The expected absolute rate of observed FRBs for each experiment, per independent tied array beam, in case of blank and lensed fields can be found in Table \ref{tab:clusters}. In the case of MACS 0717 the total rate of observed FRBs increases by $\sim20\%$ for both CHIME and CHORD-64, and nearly doubles in the case of CHORD-512. For CHORD-512 the cluster occupies a significantly greater portion of the beam than for CHORD-64 and CHIME, and therefore the effects of lensing are larger in this case, leading to its greater relative gain in total rate. In the case of $z\geq1$ FRBs, the CHIME rate increases by nearly a factor of six, CHORD-512 by three, but CHORD-64 only 1.7. As seen in Fig. \ref{fig:combined} CHIME's sensitivity in z--DM space is concentrated at $z\lesssim1$. Under the model that FRBs trace star formation rate, the number of FRB sources will increase rapidly towards cosmic noon ($z\sim2$) resulting in a larger number of sources undetected by CHIME at $z\geq1$. The relative gain in $z\geq1$ FRBs for CHIME from lensing is substantial. Conversely, both CHORD instruments have pre-existing sensitivity to high redshifts, resulting in lensing extending their range into even higher redshift domains where the star formation rate and thus number of FRB sources is greatly diminished, restricting the probability of their detection in our model, but providing an opportunity to test how well FRBs trace star formation, see discussion in \S \ref{sec:discussion}.

\begin{figure}
    \centering
    \includegraphics[width=0.8\linewidth]{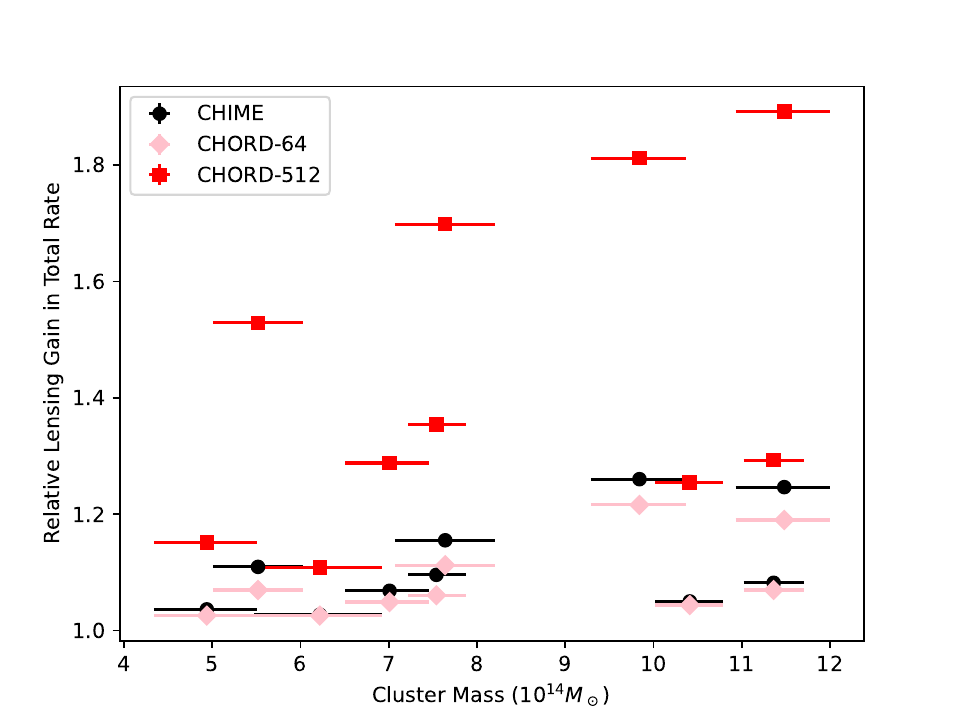}
    \includegraphics[width=0.8\linewidth]{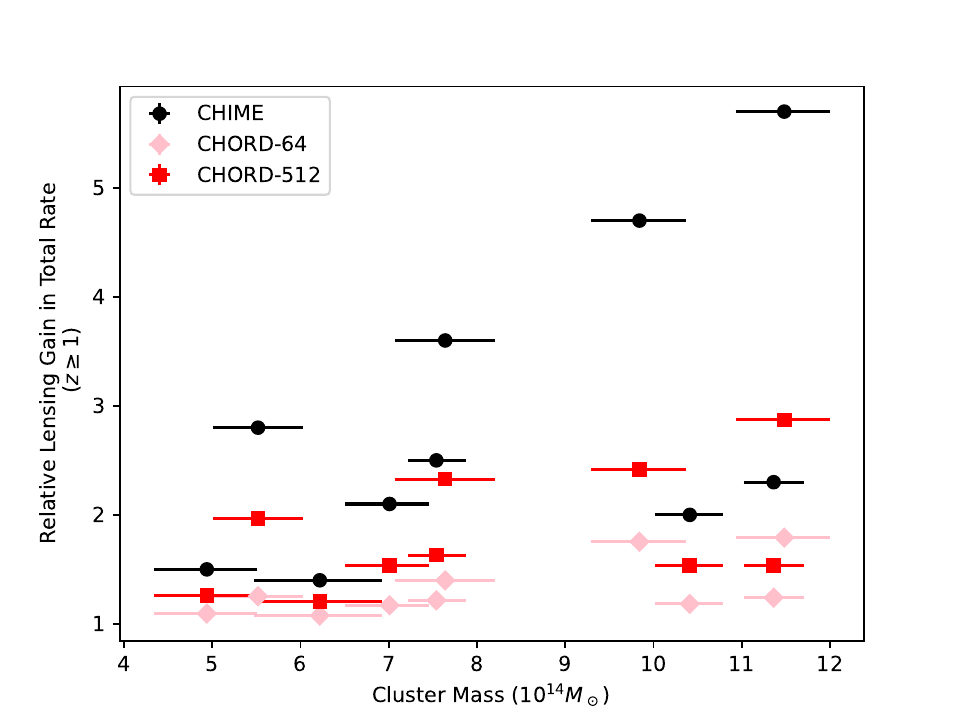}
    \caption{Relative gain in observed rates (as defined in Eq. \ref{eq:totalRate}) due to lensing as a function of cluster mass for CHIME, CHORD-64 and CHORD-512 for both total rates (\textit{top}) and rates at $z\geq1$ (\textit{bottom}).}
    \label{fig:RateGainvsMass}
\end{figure}

The rates and z--DM distributions for the remaining clusters can also be found in Table \ref{tab:clusters} and appendix \ref{app:clusters} respectively. Each shows similar behaviour to that demonstrated for MACS 0717 above, but to lesser extents. Figure \ref{fig:RateGainvsMass} shows the relative gain in observed FRB rate integrated over DMs for both all redshifts and redshifts above one as a function of the mass for each cluster in the sample and each instrument. Primarily, the Figure shows a mild correlation between cluster mass and lensing gain, however the morphology of the cluster clearly plays a significant role, contributing scatter of similar amplitude to the rate gain. This scatter suggests that when estimating the impact of lensing from a cluster it is important to account for its unique morphology, justifying the more complex modelling we perform in this study. Secondarily, the plot also shows that the relative impact of lensing between each instrument is common to all lenses, as expected for the instrumentally driven justifications given above. 

\section{Discussion}\label{sec:discussion}

From cosmology to source studies, high redshift FRBs are a valuable probe that have thus far been difficult to robustly identify. The expectations of lensing shown in \S\ref{sec:results} demonstrate that the massive galaxy cluster lenses can dramatically increase the number of high redshift FRBs observed through synthesised beams containing those clusters. The mild correlation between the mass of a cluster and the relative gain in observed burst rate seen in Fig. \ref{fig:RateGainvsMass} indicates that the largest clusters, like MACS 0717, will allow for the greatest number of high redshift bursts to be observed in any given beam. The number of extremely massive lenses like MACS 0717 is very limited however, with clusters $\geq10^{15}\,$M$_\odot$ constituting $\lesssim1\%$ of the clusters found by large surveys \citep{bleem_galaxy_2015} and often being absent from smaller surveys altogether \citep{chiu_erosita_2022}. For steerable instruments like Murriyang, Green Bank Telescope, ASKAP, MeerKAT and to some extent FAST, this small number of massive clusters can be tracked for long exposures and therefore remain useful targets for collecting high redshift FRBs. In lieu of any overriding motivation, we recommend that FRB searches on steerable instruments always track the cluster with the largest highly magnified area in its lens model. Conversely, for transit instruments such as CHIME and CHORD, the most massive clusters will only fall within the primary beam for a small window of time each day, limiting their total exposure and stifling the advantage of cluster lensing. Due to the wide field of view of both CHORD and CHIME however, we can instead appeal to the greater number of lower mass clusters and build our high redshift sample by stacking their comparatively lesser effects. 

The eROSITA Final Equatorial Depth Survey \citep[eFEDS;][]{bulbul_erosita_2022} conducted uniform x-ray observations over a 140 square degree field from which a highly complete sample of clusters ($\gtrsim 95\%$), down to low mass ($\gtrsim 10^{13}\,$M$_{\odot}$), can be extracted. This commissioning survey demonstrates the sensitivity of eROSITA in a small field, providing a robust estimate of the density of low mass clusters which will be found across the sky by the full eROSITA All-Sky Survey (eRASS) in coming years. Within the eFEDS field, approximately 6 clusters $\gtrsim 5\times10^{14}\,$M$_\odot$ (the lowest mass modelled in our sample) are expected to be present \citep{chiu_erosita_2022}. These six clusters occupy a similar range in mass ($\sim 5-7.5 \times10^{14}\,$M$_\odot$) and redshift ($z\sim 0.4-0.8$) parameter space to the six lowest mass clusters in our modelled sample and therefore we assume that our clusters form an adequate representation of the eFEDS clusters. As a result, lensing from the six lowest mass clusters in our sample may be used to roughly approximate the effect of cluster lensing in an average 140 sq. degree field of view on the sky. 

\begin{figure*}
    \centering
    \includegraphics[width=\textwidth]{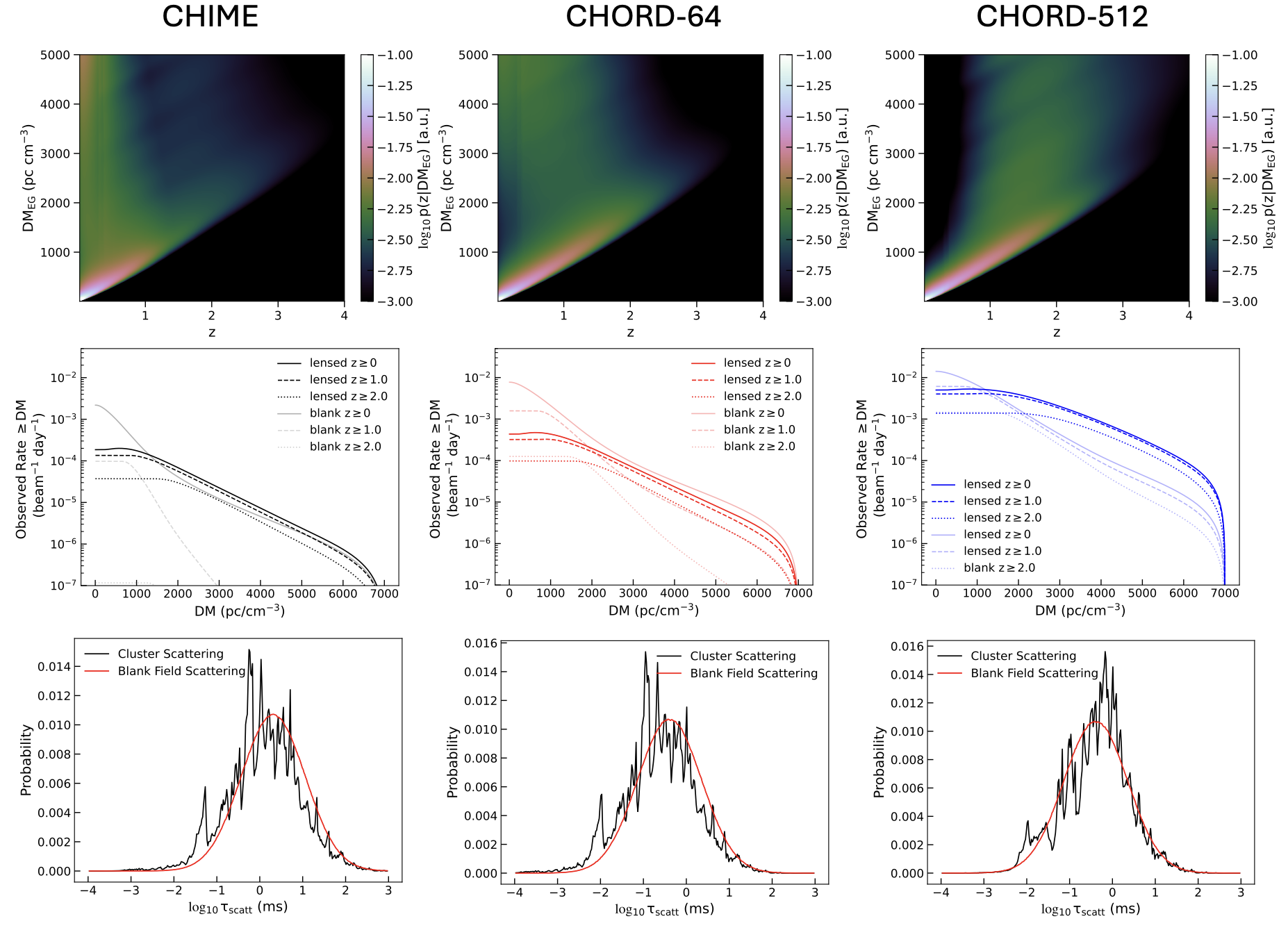}
    \caption{\textit{Top}: average z--DM distribution for low mass clusters observed by CHIME, CHORD-64 and CHORD-512 (\textit{left} to \textit{right} respectively). \textit{Middle}: Cumulative rates of bursts detected above DM for CHIME, CHORD-64 and CHORD-512 for varying redshift cutoffs. \textit{Bottom}: Scattering contribution from low mass clusters compared to the expected scattering in blank fields as fit from the CHIME catalog \citep{shin_inferring_2022}, scaled for each instrument assuming $\nu^4$ kolmogorov turbulence.}
    \label{fig:lowmass}
\end{figure*}

Averaging the z--DM distributions of the remaining six low mass clusters yields Fig \ref{fig:lowmass}. As expected the average impact of the lower mass clusters on the observed z--DM distribution is lesser than the case of MACS 0717. However, even these lower mass lenses still provide a significant advantage over the blank field case. 

Integrating over redshift, the middle row of Fig. \ref{fig:lowmass} shows the excess rates above the blank field case corresponding to each average z--DM distribution depicted in the top row of Fig. \ref{fig:lowmass}. The rates are calculated for observations above a given DM and redshift, showing both the complete and high redshift detection rates when a given DM cutoff is applied. For comparison the blank field cases of each instrument are also shown. For each line, the plateau's indicate the DM, below which the observed rate does not increase substantially. In the case of the blank fields with $z\geq0$, there is no plateau because nearby FRBs with DMs approaching zero contribute to the rate. Conversely in the lensing excess there is a plateau even in the case of $z\geq0$, as FRBs below the lens redshift do not contribute to the excess. Indeed the lensing excesses even show a downturn at intermediate DMs ($\sim500-1000\,$pc cm$^{-3}$) due to the cluster increasing the DM of those FRBs at intermediate distances. Summed cumulatively the loss of those intermediate DM sources manifests as the downturn seen in Fig. \ref{fig:lowmass}. 

For each instrument and redshift range shown in Fig. \ref{fig:lowmass}, the excess rates from lensing lie close to or above their blank field counterparts in the high DM regime (DM$\gtrsim2000\,$pc cm$^{-3}$). Broadly, this indicates that on average, cluster lensing will at least double the observed rate of high DM FRBs in blank fields, with the fraction of those bursts coming from high redshifts ($z\gtrsim1$) often jumping by orders of magnitude. These conclusions should hold for all instruments with sensitivity to observe FRBs at least as distant as the cluster in a blank field. Furthermore, the gain due to lensing will be accentuated in beams smaller than those modelled here, as the excess rate from the cluster will remain constant (provided the beam remains larger than the cluster) but the blank field rate corresponding to a smaller beam will be lower. We highlight however that instruments with smaller beams often have greater sensitivity and, as discussed in \S \ref{sec:results}, instruments with sensitivity sufficient to detect FRBs over a majority of their redshift distribution, such as CHORD-512 under the assumption that FRBs trace star formation, will see less improvement in high redshift burst rates than less sensitive instruments. In these cases cluster lenses simply serve to increase the overall rate by probing the FRB distributions at each distance to lower energies. For an instrument such as FAST, an FRB search of the close-packed beams in the multi-beam receiver would cover a sky area similar to what we model for a single CHORD-512 synthesised beam and therefore targetting a lensing cluster would be expected to increase observed rates by a factor of $\sim2-3$ compared to the rate in a blank field, depending on the cluster's precise morphology. This would yield results equivalent to 2-3 CHORD-512 beams, for a similar sensitivity.

If, contrary to expectation, the FRB redshift distribution continued steadily to redshifts beyond the peak of star formation, the rates observed through lensing clusters would be many times higher. For example, if the redshift distribution continued as $\Phi\propto(1+z)^{2.7}$, following CSFR in the nearby universe (see Eq. \ref{eq:CSFR}), the relative gain from cluster lensing to FRBs detected between redshifts 1–4 would scale from 10 to 100, respectively. The observation of lensing clusters by sensitive facilities such as CHORD-512 or FAST, therefore, provides a large lever arm with which the redshift distribution of FRBs can be constrained. Moreover, if an appreciable population of FRBs exists at very high redshifts ($z\gtrsim6$) then cluster lines of sight are likely to contain the highest density anywhere on the sky of FRBs which are detectable from the epoch of reionization. This makes galaxy cluster lenses the most natural lines of sight search for FRBs with which to probe the properties of reionization\footnote{A caveat here is that the DM contribution from the cluster must also be precisely known if these high redshift FRBs are to be useful in probing reionization.} \citep{fialkov_constraining_2016, pagano_constraining_2021}.

Distant FRBs must necessarily occur at relatively high DMs and therefore high redshift candidates are typically identified amongst unlocalised FRBs, by filtering out only those with DMs above some threshold. Using the rates per beam shown in the middle row of Fig. \ref{fig:lowmass} we can evaluate the efficacy of such DM thresholding for each telescope. We characterise the efficacy by considering three quantities of the filtered sample, the completeness, denoting the fraction of the total high redshift sample that remains,
\begin{align}\label{eq:completeness}
    \text{Completeness} &= \frac{(\text{DM}>\text{DM}_T)\&(z\gtrsim z_\text{lim})}{(z\gtrsim z_\text{lim})},
    \intertext{the purity, denoting the fraction of the filtered sample which are high redshift,}
    \text{Purity} &= \frac{(\text{DM}>\text{DM}_T)\&(z\gtrsim z_\text{lim})}{(\text{DM}>\text{DM}_T)},
\end{align}
and the size/absolute rate of that sample. For each telescope, the observed rate towards low DMs in any beam is dominated by $z\lesssim1$ bursts observed from the blank field component, and therefore DM thresholds (DM$_T$) in the range DM$_T\sim0-1000\,$pc cm$^{-3}$ are of little value for isolating high redshift bursts. For blank fields in Fig. \ref{fig:lowmass} the fraction of high redshift bursts is often highest near the edge of that redshifts cumulative rate (i.e. ratio of $z\geq1$ / $z\geq0$ is greatest at the edge of the cumulative $z\geq1$ plot), with that fraction falling precipitously towards higher DMs where bursts are more likely to occur nearby with large host DMs, consistent with the results of \cite{james_z--dm_2021}). Conversely the excess rates due to lensing have shallower gradients with DM$_T$ even for the high redshift components, and therefore increasing DM$_T$ can improve the fraction of observed bursts at high redshift, i.e. host DM contributions become washed out by cluster DMs. Of course, higher DM$_T$ values also result in fewer observed bursts. In each case we must therefore select DM$_T$ to optimise both the high redshift fraction, and rate of observed bursts. The amount to which fraction is favoured over rate in this optimisation is arbitrary and therefore we focus our discussion on rates which optimise both when weighted equally. 

For CHIME, CHORD-64 and CHORD-512 the DM$_T$ maximising both high redshift completeness and purity for the blank and lensed fields cases at various redshift cut-offs can be found in Tables \ref{tab:DMT2} \& \ref{tab:DMT2}. As expected the higher redshift conditions require larger DM$_T$s, with CHIME showing substantially lower DM$_T$ values (especially at $z\gtrsim2$) due to its limited sensitivity as seen in Fig. \ref{fig:combined}. In the cluster lensed case the optimal DM$_T$ is marginally larger than for blank beams for all telescopes, due to the excess DM contributed by the cluster. Notably the increase in the optimal DM$_T$ is substantially less than the expected contribution from the cluster $\mathcal{O}(10^3)\,$pc cm$^{-3}$, likely to preserve the high redshift contributions from the unlensed parts of the beam. The completeness of all filters is $\gtrsim90\%$ indicating that typically less than $\sim10\%$ of the potentially detectable high redshift events are discarded by the DM filtering in either the blank or lensed cases. 

The largest impact of lensing manifests differently for telescopes of different sensitivity. In the case of CHIME, the survey sees a substantial boost in both the purity and rate of high redshift events within the filtered sample, with each doubling for $z\geq1$ and increasing by several orders of magnitude for $z\geq2$. Whereas in the case of either CHORD configuration the impact on purity is curtailed by the high pre-existing purity in blank fields and therefore the impact of lensing is limited to the increases in absolute rate. For CHORD-64 this can be seen in the $z\geq1$ bursts, where the improvement to purity is only $5\%$ but the improvement to rate is higher at $20\%$, whereas for the more distant $z\geq2$ bursts, where blank field purity is low, the boost to purity and rate is greater at $40\%$ and $80\%$ respectively. Similarly, CHORD-512 shows drastic improvements to absolute rate in both redshift bins, with only a small gain in purity for $z\geq1$ and a significant drop in the purity at $z\geq2$ in the lensed beam case. This decrease in purity results from the qualitatively different shape of the lensed z--DM distribution for CHORD-512, as shown in Fig. \ref{fig:lowmass}. Unlike CHIME and CHORD-64 where the contributions from lensing are similar or sub-dominant to the blank field background (see the middle row of Fig. \ref{fig:lowmass}) and therefore only marginally shift the DM distribution, CHORD-512 is dominated by the lensing rate at large DMs due to the cluster occupying a substantially larger portion of the beam and therefore typical redshift DM relations become washed out by the cluster contribution at all redshifts beyond the mean cluster redshift ($z\sim0.4$), causing the decrease in purity towards higher redshifts compared to the blank fields case. The primary advantage of lensing for sensitive instruments is therefore the relative gain in total rate. Below we consider the total rates corresponding to each array configuration for various observational scenarios, and discuss the value of lenses in gathering high redshift samples in each case.

\begin{table*}
\centering
\setlength{\arrayrulewidth}{0.5mm} 
\renewcommand{\arraystretch}{1.5} 

\begin{tabular}{|c||c|c|c|c|}
\hline
\textbf{Survey} & \multicolumn{4}{c|}{\textbf{Blank}} \\ \hline
& DM$_T$ (pc cm$^{-3}$) & Completeness & Purity & Rate (1000 Beam Days)$^{-1}$ \\ \hline
CHIME ($z\gtrsim1$)     & 790  & 0.91 & 0.18 & 0.1\\ \hline
CHORD-64 ($z\gtrsim1$)     & 850  & 0.92 & 0.53 & 1.58\\ \hline
CHORD-512 ($z\gtrsim1$) & 860  & 0.95 & 0.75 & 6.13\\ \hline
CHIME ($z\gtrsim2$)     & 1370 & 0.96 & 0.00 & 0.00\\ \hline
CHORD-64 ($z\gtrsim2$)     & 1630 & 0.89 & 0.17 & 0.13\\ \hline
CHORD-512 ($z\gtrsim2$) & 1700 & 0.89 & 0.50 & 1.46\\ \hline
\hline
& \multicolumn{4}{c|}{\textbf{Lensed}} \\ \hline
 & DM$_T$ (pc cm$^{-3}$) & Completeness & Purity & Rate (1000 Beam Days)$^{-1}$ \\ \hline
CHIME ($z\gtrsim1$)     & 900  & 0.89 & 0.37 & 0.23 \\ \hline
CHORD-64 ($z\gtrsim1$)     & 870  & 0.92 & 0.56 & 1.9\\ \hline
CHORD-512 ($z\gtrsim1$) & 890  & 0.96 & 0.77 & 10.1\\ \hline
CHIME ($z\gtrsim2$)     & 1880 & 0.87 & 0.25 & 0.04\\ \hline
CHORD-64 ($z\gtrsim2$)     & 1700 & 0.89 & 0.24 & 0.23\\ \hline
CHORD-512 ($z\gtrsim2$) & 1750 & 0.94 & 0.39 & 2.84\\ \hline
\hline
& \multicolumn{4}{c|}{\textbf{Lensed -- Morphology Filtered}} \\ \hline
                  & DM$_T$ (pc cm$^{-3}$) & Completeness& Purity& Rate (1000 beam days)$^{-1}$ \\ \hline
CHIME ($z\gtrsim1$)            & 975 & 0.63 & 0.58& 0.15\\ \hline
CHORD-64 ($z\gtrsim1$)            & 905 & 0.35 & 0.62& 0.67\\ \hline
CHORD-512 ($z\gtrsim1$)        & 990 & 0.54 & 0.78& 5.43\\ \hline
CHIME ($z\gtrsim2$)            & 1835 & 0.89 & 0.35& 0.03\\ \hline
CHORD-64 ($z\gtrsim2$)            & 1780 & 0.52 & 0.31& 0.11\\ \hline
CHORD-512 ($z\gtrsim2$)        & 2125 & 0.55 & 0.39& 1.56\\ \hline
\end{tabular}
\caption{Rates and efficacy parameters for bursts above DM$_T$ for CHIME, CHORD-64 and CHORD-512 surveys at various redshift targets for blank fields, cluster lensed fields and morphology filtered cluster lensed fields.}
\label{tab:DMT2}
\end{table*}



\subsection{Blind Survey}\label{subsec:blindSurvey}
While large cluster surveys like the eFEDS and the SPT survey constrain well the number count of galaxy clusters on the sky, these surveys only cover a relatively low fraction of the total sky area. As a result, the most likely scenario for cluster lensing in FRB surveys now is one where the expected number of clusters within the primary beam's field of view is known, but their precise locations are undetermined. In such a case the localised position of an FRB is irrelevant for determining whether it passes through a cluster and we must rely on DM alone to isolate our high redshift sample\footnote{We note that, given the increased rate from cluster lines of sight, methods that use FRB localisations and DM to constrain the probability of chance coincidence on the sky, such as \cite{cook_k-contact_2024}, which has been applied to identify previously unknown repeating FRBs, may also be applied to determine the position of previously unknown clusters. This method could be applied in combination with the Planck Comptop-Y map to constrain the all sky clustering of FRBs with with massive galaxy clusters}. 

While clusters can clearly contribute many more high redshift bursts than blank fields, these bursts are mixed with many high DM$_\text{Host}$ bursts from nearby sources in blank fields, as shown by the substantial blank field rates at high DM despite the fall off of the higher redshift components. The relative scarcity of even lower mass lensing clusters on the sky leads to a low number of beams containing clusters compared to blank beams for widefield instruments such as CHIME and CHORD. As a direct result, when isolating a sample via DM alone, the high redshift bursts from clusters become indiscernible amongst an overwhelming number of high DM nearby FRBs from blank field beams. Fortunately, instruments such as CHIME and CHORD are designed to capture raw voltage information for bright bursts, allowing the reconstruction of burst profiles at high time and frequency resolution. This allows burst properties such as scattering timescale and scintillation bandwidth to be used in addition to DM to assess the probability observed bursts arise from lensed, high redshift candidates. 

The scattering contribution expected from clusters is shown in the bottom row of Fig. \ref{fig:lowmass} as calculated in \S \ref{sec:method}. We also plot the expected distribution of scattering times for blank fields informed by previous estimates of FRB population scattering from the first CHIME/FRB catalogue where an empirical log-normal distribution is assumed, fit with a log-normal mean of $\log\mu_s=0.7$ and standard deviation $\log \sigma_s=1.72$. We highlight that this empirical distribution changes between CHIME and CHORD due to the frequency depedence of scattering and their distinct observing bandwidths. The mean probability density function of scattering from clusters shows strong fluctuations owing to the small number of clusters in our sample, which allow physical fluctuations of each to dominate, namely in the radial density distributions shown in Appendix \ref{app:clusters}, combined with fluctuations in the lens magnification profile which weights the probability of observing FRBs at a given radius. Regardless, in the case of each telescope, the distributions of scattering expected in blank fields is comparable to the distribution of scattering contributions from clusters as seen in the bottom of Fig. \ref{fig:lowmass}. This means that a lensed FRB will on average accrue two separate scattering contributions of similar scale. The first being the blank field expectation which is common to both cases, and is thought to arise from the host galaxy or local environment \citep{chawla_modeling_2022, ocker_large_2022}, and the second from the intervening cluster. 

\begin{figure}
    \centering
    \includegraphics[width=0.8\linewidth]{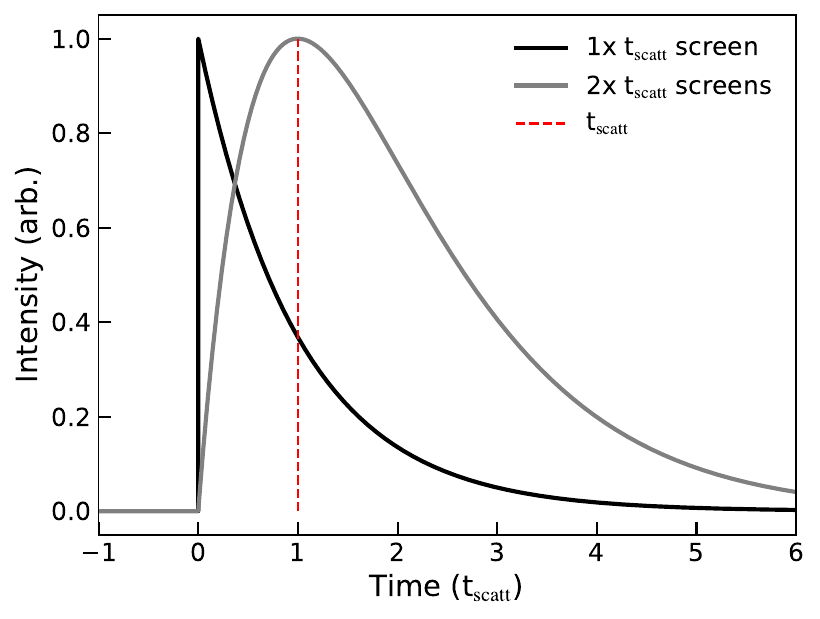}
    \caption{Impulse response functions corresponding to one and two thin scattering screens of equal characteristic scattering time.}
    \label{fig:convolvedProfile}
\end{figure}

As shown in \cite{kirsten_probing_2019}, bursts observed to pass through multiple scattering screens will be broadened following an impulse response function which is the convolution of two typical scattering profiles (sharp onset followed by an exponential tail). In such a case as here, where the scattering timescales contributed by host galaxies and clusters are comparable, this results in bursts with a rise time comparable to the scattering time from clusters and a broader tail than expected for a thin screen as seen in Fig. \ref{fig:convolvedProfile}, features which will both sharpen at higher frequencies. Such morphology provides a distinct feature which can be used to filter bursts for lensed candidates providing an advantage to telescopes with a large dynamic range in bandwidth. As a fiducial example, recent re-analysis of first CHIME FRB catalogue with the raw voltage data demonstrates that nearly $50\%$ of the observed population\footnote{Note that the triggering criteria for voltage data is typically higher due to a larger corresponding data rate and therefore the observed population in this case is only those bursts bright enough to have voltage data saved. This is a potential source of bias that we do not address here.} have at least one sub-component of intrinsic width less than 0.2\,ms \citep{sand_morphology_2024}. These bursts are therefore not expected to be from cluster lenses which as shown in Fig. \ref{fig:lowmass}, likely contribute scattering greater than 0.2\,ms to CHIME observations, resulting in rise times and inferred intrinsic widths greater than 0.2\,ms. 

The scattering contributed from clusters also results in relatively large amounts of angular broadening, particularly for high redshift FRBs, owing to the large geometric leverage associated with scattering far from the source. This large angular broadening acts to suppress scintillation in the Milky Way's interstellar medium, making it unlikely for cluster-lensed FRBs to scintillate. As a fiducial example, we can evaluate the expected index of modulation for Galactic scintillation $m_g$ in the case of scattering from a cluster, using the below equation from \cite{sammons_two-screen_2023}
\begin{equation}\label{eq:generalTwoScreen}
        m_g^2 = \frac{1}{(2\pi\nu)^2t_{\text{scatt},g}t_{\text{scatt},x}(1+z_x)}\frac{D_{s,x}D_{d,x}}{D_{ds,x}}\frac{D_{s,g}}{D_{ds,g}D_{d,g}},
\end{equation}
where $t_{\text{scatt}}$ is a scattering time, $\nu$ is the central frequency of the observation, $D_{s}$, $D_d$ and $D_{ds}$ are the angular diameter distances between observer and source, observer and scattering screen and scattering screen and source respectively, $z$ is redshift and the $x$ and $g$ subscripts denote the Galactic and cluster scattering screens. For a canonical Galactic scattering case of $t_{\text{scatt,g}}=1\,\mu s$ at a distance of $D_d=1\,$kpc, with a background cluster at a redshift of $z_x=0.4$ and an FRB source at $z_s\geq0.5$, the spectral scintillation manifested from the Galactic scattering will have modulation indices $m_g < 0.1$ for all cluster scattering times $t_{\text{scatt,x}}\geq10^{-1.5}\,$ms for each instrument considered. As both redshift and cluster scattering contribution for cluster-lensed FRBs are dominantly greater than $z_s=0.5$ and $t_{\text{scatt,x}}\geq10^{-1.5}\,$ms, we therefore expect a vast majority of FRBs passing through clusters not to exhibit Galactic scintillation. The fraction of bursts within the FRB population exhibiting Galactic scintillation is currently poorly constrained, but from the relatively unbiased sample in \cite{sammons_two-screen_2023}, four out of nine bursts were confirmed to not scintillate. While small, this sample shows that filtering scintillating FRBs out of an observed sample can potentially remove $\sim50\%$ of the bursts, allowing us to further isolate a lensed, high redshift sample.

To evaluate the expected rate of lensed FRBs over the full array we must now account for how many synthesised beams across the primary beam are searched for FRBs. As described in \S \ref{sec:method}, CHIME has 1024 tied array beams of 21.5' FWHM, as a result it is sensitive to a sky area of $\approx103$ sq. deg at the central frequency of 600\,MHz and should contain 4.4 clusters above our mass threshold on average\footnote{This is somewhat an underestimation of the true CHIME sky coverage owing to the fact that we assume circularly symmetric beam responses for the synthesised beams which is not true in practice. In order for self consistency we will continue with this assumption and apply it to each telescope similarly, as it results in an underestimation of the impact of clusters, our results remain conservative.}. This gives CHIME 1024 blank beams and 4.4 cluster beams (synthesised beams containing a cluster) on average, as each cluster is far smaller than a single beam. Due to processing power constraints, the primary beam of the CHIME telescope is only sparsely sampled by the 1024 FFT formed beams through which FRBs are detected. Given the low number of clusters within the primary beam ($\lesssim10$ at any moment), exposure to these clusters could be improved by forming tracking tied array beams to follow each cluster on the sky, likely resulting in a factor of two improvement in cluster beam days recorded for CHIME. CHORD-64 has a smaller sky coverage at its central frequency of 8 sq. deg, yielding 63.9 blank beams and an average of 0.34 cluster beams. CHORD-512 has the same field of view and therefore the same number of cluster beams, but 510.8 blank beams, due to their reduced size which is still larger than even the most extended clusters. In the case of CHORD, no additional exposure is afforded by tracking beams due to the primary beam being densely sampled by FFT formed detection beams.

Summing our rates per synthesised beam calculated in \S \ref{sec:results} over the total number of synthesised beams in each instrument we can then compare the total blank field rates with those in our filtered sample. For CHIME, using a DM threshold of 900 pc cm$^{-3}$ and filtering out 75$\%$ of the blank field population with cuts to inferred intrinsic width and scintillation behaviour creates a subsample wherein $~1\%$ of bursts pass through a Galaxy cluster lens. Given the absolute rates we calculate, CHIME should have observed at least one such cluster lensed burst after their $\sim4$ years of continuous exposure, with that burst having an $\approx70\%$ chance of coming from $z\geq1$ and a $\approx20\%$ chance of coming form $z\geq2$. The remaining $\approx100$ bursts which satisfy the above criteria but are from blank fields, come dominantly from nearby, with $\approx30$ at $z\geq1$ and none $z\geq2$.

In the case of CHORD-64, as a temporary instrument with an expected lifetime $\lesssim1$ year, we predict that the mean number of observed cluster lensed FRBs over the duration will only be $\approx0.1$, and therefore it is unlikely that CHORD-64 will detect a cluster lensed FRB using a traditional FRB survey strategy, and thus we do not consider it further until \S \ref{subsec:dynamicThresholding}. For the full CHORD-512, using a DM threshold of 890 pc cm$^{-3}$ with similar inferred intrinsic width and scintillation cuts yields a sample where $0.2\%$ passed through a cluster. Of those bursts, $77\%$ occur at $z\gtrsim1$, $26\%$ at $z\gtrsim2$. With CHORD-512 enhanced sensitivity, approximately one lensed burst is expected to be observed for every year of continuous exposure. With the blind survey strategy this one lensed burst per year will be hidden amongst $\approx350$ bursts from blank fields with $270$ at $z\gtrsim1.0$, $70$ at $z\gtrsim2$ and $10$ at $z\gtrsim3$. 

Overall for blind surveys, where the position of clusters is unknown, lenses contribute $\lesssim1-2\%$ of high DM, non-scintillating FRBs with inferred intrinsic widths greater than 0.2\,ms. For CHIME, these lensed bursts are hidden amongst contaminating bursts from the nearby universe with high host DMs, and in the case of CHORD-512 the lensed bursts are mixed with other genuinely high redshift FRBs. In either case, while the criteria we outline do provide a way to define a sample of FRBs with a slightly higher concentration of high redshift events, the improvement afforded by considering lensing, in the context of a blind survey, is slight owing to the effects of cluster lensing being washed out by the larger number of blank field beams. As such we can conclude that to make effective use of clusters, their positions must be known \textit{a priori}.

\subsection{Cluster Positions Known}
In the case where an FRB survey shares substantial overlap with known cluster surveys, or in the eventuality where planned x-ray programs like eROSITA's All Sky Survey (eRASS) have been completed, many clusters may be known. Taking a subsample of FRBs collected only in beams containing clusters renders a population described by Fig. \ref{fig:lowmass} and Table \ref{tab:DMT2}. The purity of the high redshift samples described in Table \ref{tab:DMT2} can be improved even further, at the cost of completeness, by performing the morphology cuts outlined above on the cluster beams prior to calculating the optimum DM$_T$. The associated completeness and purity of the morphology filtered, lensed samples are shown in Table \ref{tab:DMT2}. We do not update the blank field cases from those in Table \ref{tab:DMT2} as intrinsic width and scintillation should be at most only weakly redshift dependent, and therefore any changes to completeness and purity will be second order effects. For CHIME there is a substantial improvement to the purity of high redshift bursts in the filtered sample, with a commensurate drop in completeness. Conversely for both CHORD arrays the improvement to purity is small to negligible at the cost of a large drop in completeness and therefore rate. This occurs due to a large number of genuinely high redshift events from blank fields being filtered out by our morphology selection criteria. We therefore recommend that when aiming to isolate a significant number of high redshift bursts in a sample of high purity that cluster beam samples from CHIME should be filtered on DM and burst morphology, whereas cluster beam samples from CHORD should be filtered on DM alone. 



From the rates displayed in Table \ref{tab:DMT2}, the sample collected by CHIME over four years will contain on average, 1.6 bursts, with 0.9 at $z\geq1$ for DM$_T\geq975\,$pc cm$^{-3}$ or 0.6 bursts with 0.2 at $z\geq2$ for DM$_T\geq1835\,$pc cm$^{-3}$. From Table \ref{tab:DMT2} (i.e. no morphology filtering) the CHORD-64 rate is once more too low to be significant, with only $\sim120$ beam days on clusters over its lifetime. Conversely, CHORD-512 is expected to capture on average 1.6 events per year, 1.3 of which occur at $z\geq1.0$, for DM$_T\geq890\,$pc cm$^{-3}$ and 0.9 events per year, 0.3 of which occur at $z\geq2.0$, for DM$_T\geq1750\,$pc cm$^{-3}$. Compared to the blind survey case discussed in \S \ref{subsec:blindSurvey}, the morphology and cluster position filtered samples have far fewer unlensed FRBs, improving purity of high redshift FRBs in the final filtered samples. Each of the filtered samples (except for CHORD-512 at z$\geq2.0$ for reasons outline at the beginning of \S \ref{sec:discussion}) now have a purity of high redshift FRBs greater than the blank field cases described in Table \ref{tab:DMT2}. This improvement to purity is of significant utility as it can improve our ability to conduct high redshift studies independently of individual redshift determinations. Furthermore it can allow follow-up resources to be dedicated to high redshift FRB targets with greater confidence, ultimately resulting in a greater number of confirmed high redshift hosts. For simplicity we have considered here only the case where the cluster positions are known definitively, however the same survey could also be informed using a probabilistic distribution of cluster positions (e.g. making use of marginal cluster detections) by forming a joint probability distribution with the lensed z--DM distributions derived here for the average low mass cluster.

While the likelihood of high redshift events within an observed sample has been greatly improved, the size of the resulting samples is still small however, owing to the limited number of even the lower mass galaxy cluster lenses on the sky. The number of FRBs within these high purity lensed samples can be boosted electronically, by focussing a telescopes backend processing power on known cluster lenses through a process we dub \textit{dynamic thresholding}.

\subsection{Improving the Advantage with Dynamic Thresholding}\label{subsec:dynamicThresholding}
For large scale FRB survey programs like CHIME, and in the future CHORD, the ability to detect FRBs is limited not by the sensitivity of the instrument, but by the number of false positive candidates. Within CHIME the detection threshold for FRBs is set to a S/N$\approx8$. Certainly FRBs at lower S/N values may still be distinguished from noise sufficiently well to be detected. However, towards lower S/N thresholds the number of false positive signals contributed by non-gaussian contaminants like RFI becomes overwhelmingly large. Under the assumption of homogeneity each synthesised beam searched for FRB signals is treated equally, with a constant detection S/N (equivalently fluence) for all beams set by the maximum number of true and false candidates which can be sorted with the available resources. 

In this paradigm the increase in candidate events prohibits lower fluence thresholds from being applied to all beams. By increasing the fluence threshold for the majority of beams a small number of beams could be searched to lower fluence thresholds while maintaining a constant number of candidates collected over the entire array. As clusters are expected to occupy only a few beams within a given field of view, they are uniquely suited as targets of a revised search strategy that dynamically changes the detection fluence threshold in beams containing a cluster. Such a strategy is already employed by CHIME/FRB to detect faint bursts from lines of sight containing repeating FRBs and could increase the number of FRBs observed from cluster lines of sight, expanding the size of the statistically high redshift samples associated with them.

\begin{figure}
    \centering
    \includegraphics[width=0.8\linewidth]{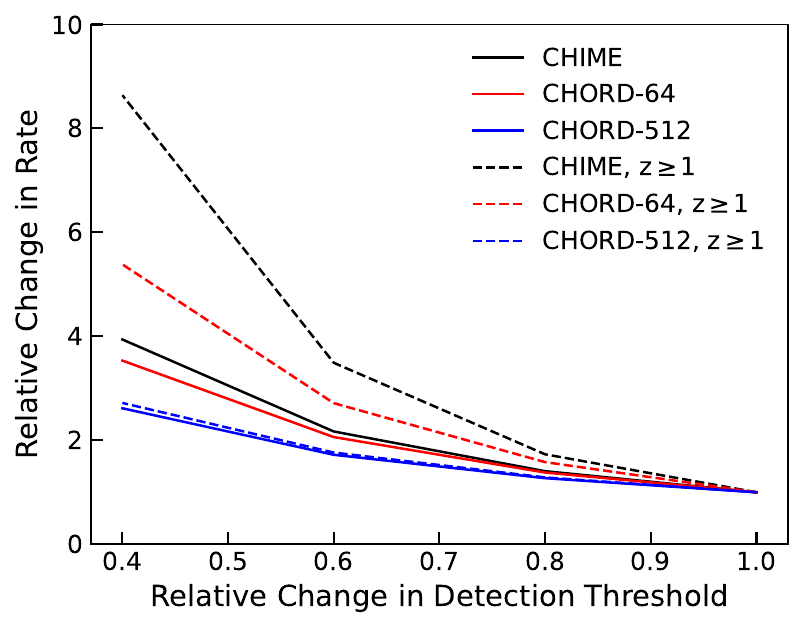}
    \caption{Relative boost in detected rates for all redshifts (full) and redshifts $z\geq1$ (dashed) as a function of relative change in detection threshold for CHIME, CHORD-64 and CHORD-512.}
    \label{fig:dynamicthreshboost}
\end{figure}

The amount to which these samples are increased, depends on how low the detection threshold is dropped. Fig. \ref{fig:dynamicthreshboost} shows the fractional change in observed rates from cluster beams for all bursts and those at $z\geq1$, with only small changes to purity noted in the lower fluence samples\footnote{Towards lower fluence thresholds distant FRBs can be detected more readily and hence the purity of collected samples increases slightly.}. By dropping the detection threshold to 80\% of its previous value the total number of detected bursts increases for all telescopes by approximately 50\%. Towards even lower fluence the improvement in rate diverges for different telescopes, with CHIME and CHORD-64 increase close to the euclidean expectation of $R\propto F_0^{-1.5}$ as they continues to detect more distant bursts. Conversely, the CHORD-512 improves more slowly at rate close to $R\propto F_0^{\gamma}$ as both detectability and star formation rate at higher distances becomes suppressed. Looking instead at only the change in rates at $z\geq1$ the relative change in rate is greater due to the majority of all bursts at higher fluence being contained at lower distances. To determine how low the fluence of a given beam can be dropped we must better understand how the number of false positive events introduced by RFI and other non-gaussian noise sources changes as a function of detection threshold. At present this function is not well understood, but will be the focus of a future work. For now we simply conclude that cutting the detection threshold to 60\% (e.g. from S/N 8 to 5 in the case of CHIME) through dynamic thresholding, would approximately double the total number of bursts observed through cluster lines of sight, and would increase the rate of $z\geq1$ bursts listed in Table \ref{tab:DMT2} by factors of $\approx$2--4. In such a scenario both CHIME and CHORD-64 could collect on average approximately 1.5 bursts every year from cluster lines of sight with DM$\geq$DM$_T(z\geq1)$ from Tables \ref{tab:DMT2} \& \ref{tab:DMT2} respectively, with a purity above 50\% in each case. For CHIME this represents a substantial advantage, where currently the same number of beams directed at blank fields collect fewer than one FRBs per year at DM$\geq$DM$_T(z\geq1)$ (Table \ref{tab:DMT2}), with a purity below 20$\%$. For CHORD-64 the benefit is similar with the same cluster beams detecting fewer than one FRBs per year above DM$_T(z\geq1)$, with a purity of $\approx50\%$, for a default search strategy on a blank field. For CHORD-512 dynamic thresholding would allow 3 bursts per year above DM$_T(z\geq1)$ to be detected, with a purity $\approx75\%$, and would allow 1.5 bursts with DM$_T(z\geq2)$ to be detected, with a purity of $\approx40\%$ (i.e. an average yearly rate of 0.6 for $z\geq2.0$). In comparison the same beams in blank fields with a default search strategy would detect only one burst above DM$_T(z\geq1)$, with a similar purity of $\approx75\%$ and are unlikely to detect any at $z\geq2$ in a single year. 

\section{Conclusion}\label{sec:conclusion}

The gravitational lensing from galaxy clusters has proven to be a useful tool for observing distant sources in many fields of astrophysical research. As we've shown this utility can also benefit FRBs, with cluster lenses expected to provide a significant boost to both the rate and range of FRB observations made with current and future instruments. By considering the impact of clusters on CHIME, CHORD-64 and CHORD-512 for a range of observing scenarios we have determined how best to leverage cluster lensing to enhance the samples of valuable high redshift FRBs, even in the cases where host associations are impossible. When planning future observations the primary takeaways of this study are:

\begin{itemize}
    \item The presence of a low mass cluster ($M_\text{Planck}\geq10^{14}\,M_\odot$) in a detection beam will on average approximately double the rate of $z\geq1$ FRBs detected by CHIME and $z\geq2$ FRBs detected by CHORD-64 or CHORD-512, as seen in tables \ref{tab:DMT2}.
    \item Instruments with insufficient sensitivity to detect FRBs beyond the peak of cosmic star formation $z\sim2$ (such as CHIME) receive a significant increase to both the range of redshifts over which FRBs are detected and the total rate of observed FRBs when a cluster is present within a detection beam with FWHM $\lesssim 20$ arcminutes.\\
    \item Instruments with sufficient sensitivity to detect FRBs over redshifts containing the vast majority of star formation ($z\lesssim4$, such as CHORD) receive a diminishing increase in the range of detected redshifts when a cluster is present within a detection beam, but still benefit from an increase total detection rate.\\
    \item Observations of lensing clusters by instruments with sufficient sensitivity to detect FRBs over redshifts containing the vast majority of star formation provide a long lever arm for constraining high redshift source densities, allowing even non-detections to strongly constrain the redshift distribution of FRBs. Furthermore, such observations provide the best chance at constraining the epoch of reionization with FRBs.\\
    \item The relative gain in observed high redshift FRBs is correlated with cluster mass but with a near unity variance based on cluster morphology. Therefore to optimise the effect of lensing on observed samples, steerable instruments should track the largest available cluster with a known lensing profile. Whereas, transit telescopes should marginalise over cluster morphology by tracking a larger number of low mass clusters, due to low incidence of high mass clusters in a transiting field of view.\\
    \item Lower mass lenses are still scarce on the sky, with a density of $\sim4$ per 100 sq. degrees for $M_\text{lens}\gtrsim5\times10^{14}\,M_\odot$. As a result of this scarcity, the effect of lensing is almost completely washed out when averaged blindly over the whole sky, i.e. cluster positions must be known for the effect of lensing to become identifiable.\\
    \item When cluster positions are known the samples collected from cluster beams can be used to construct samples with a high purity of high redshift bursts. Further downsampling based on burst morphologies expected from clusters can improve the purity of these samples. For both CHIME and CHORD we find that filtering out all bursts below a threshold DM$_T \sim (1000\times z)\,$pc cm$^{-3}$ (specific values can be found in tables \ref{tab:DMT2}, \ref{tab:DMT2} and \ref{tab:DMT2}) markedly improves high redshift purity. Furthermore for CHIME we find that purity can be improved further still, without significantly compromising completeness by filtering out all bursts displaying either Galactic spectral scintillation or a sub-component width below 0.2\,ms.  \\
    \item To optimise the effect of lensing, transit telescopes should treat detection beams containing clusters differently to blank sky beams. We advocate here for a dynamic thresholding approach where cluster beams have a lower detection threshold. \\
    \item Using the above recommendations and filters (cluster beams, DM$_T$ and morphology) we expect CHIME and CHORD-64 to collect at least one burst a year above $S/N=5$, with 50\% coming from $z\gtrsim1$. This is a $100\%$ increase in the rate expected from the same beams on blank fields, and a $250\%$ increase in purity for CHIME. For CHORD-512 three bursts at $z\geq1$ and 1.5 at $z\geq2$ per year, more than doubling the blank beam rate. In each case the vast majority of such bursts would also be gravitationally lensed.\\
    \item For transit telescopes with a sparsely sampled primary beam such as CHIME, tracking beams should be used to enhance cluster exposure. We expect that the above calculated rates would double as a result to four bursts per year at $z\geq1$. 
\end{itemize}

\section*{Acknowledgements}
We thank Clancy W. James for helpful instructions on working with the z--DM repository and Robert Main for visualisation insights. M.W.S. acknowledges support from the Trottier Space Institute Fellowship program. M.D. is supported by a CRC Chair, NSERC Discovery Grant, and CIFAR. Some/all of the data presented in this paper were obtained from the Mikulski Archive for Space Telescopes (MAST). STScI is operated by the Association of Universities for Research in Astronomy, Inc., under NASA contract NAS5-26555. Support for MAST for non-HST data is provided by the NASA Office of Space Science via grant NNX13AC07G and by other grants and contracts.

\section*{Data Availability}

The data underlying this article will be shared upon reasonable request to the corresponding author.



\bibliographystyle{mnras}
\bibliography{references}



\appendix

\section{3D Electron Distribution Model of MACS J0717.5+3745}\label{app:3DEDModel}
\begin{figure}
    \centering
    \includegraphics[width=0.8\linewidth]{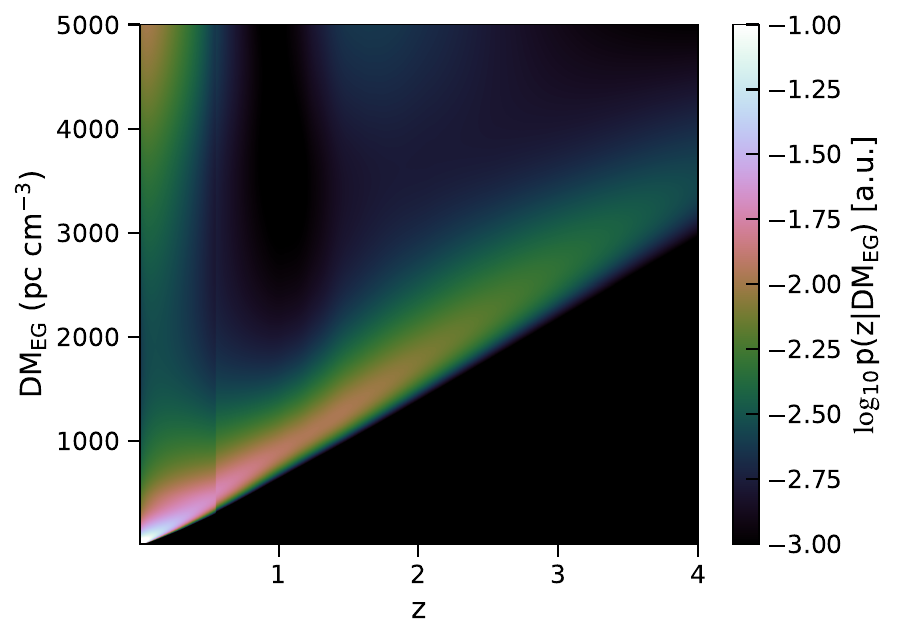}
    \caption{z--DM distribution expected to be observed in a synthesised beam of the CHIME telescope through MACS J0717.5+3745 for the case of the three dimensional, four sub-cluster electron distribution model of \citet{breuer_thermodynamic_2025}}
    \label{fig:non-radial}
\end{figure}

Throughout this work we make the simplifying assumption of a uniform cluster depth of 1\,Mpc. Here we examine the level of uncertainty introduced by this assumption into the lensed z--DM distributions by comparing to a more complex, three dimensional electron distribution model for MACS J0717.5+3745. We use the four sub-cluster model employed by \cite{breuer_thermodynamic_2025} to describe the projected electron distribution and then calculate the z--DM distribution for the CHIME telescope as described in \S \ref{sec:method}. Fig. \ref{fig:non-radial} shows the resulting z--DM distribution which shows up to a factor of two change in rates marginalised over redshift, when compared with the uniform depth results shown in Fig. \ref{fig:combined}, but a less than 10\% change in rates marginalised over DM. We therefore conclude that using a simplistic model of cluster electron distributions has little impact on the expected ability of cluster lenses to probe high redshifts, but presents a source of significant uncertainty in the expected DM distribution of lensed bursts.

\section{Cluster Morphology and z--DM}\label{app:clusters}

\begin{figure*}
    \centering
    \includegraphics[width=0.6\linewidth]{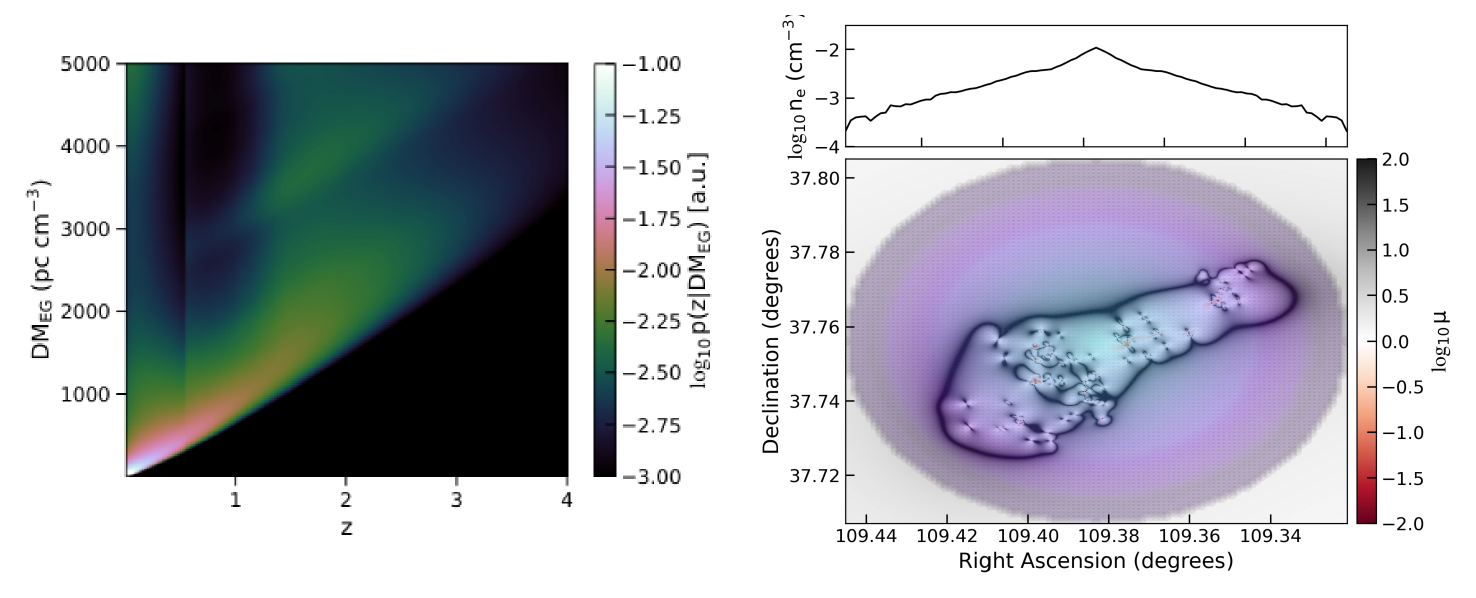}
    \includegraphics[width=0.6\linewidth]{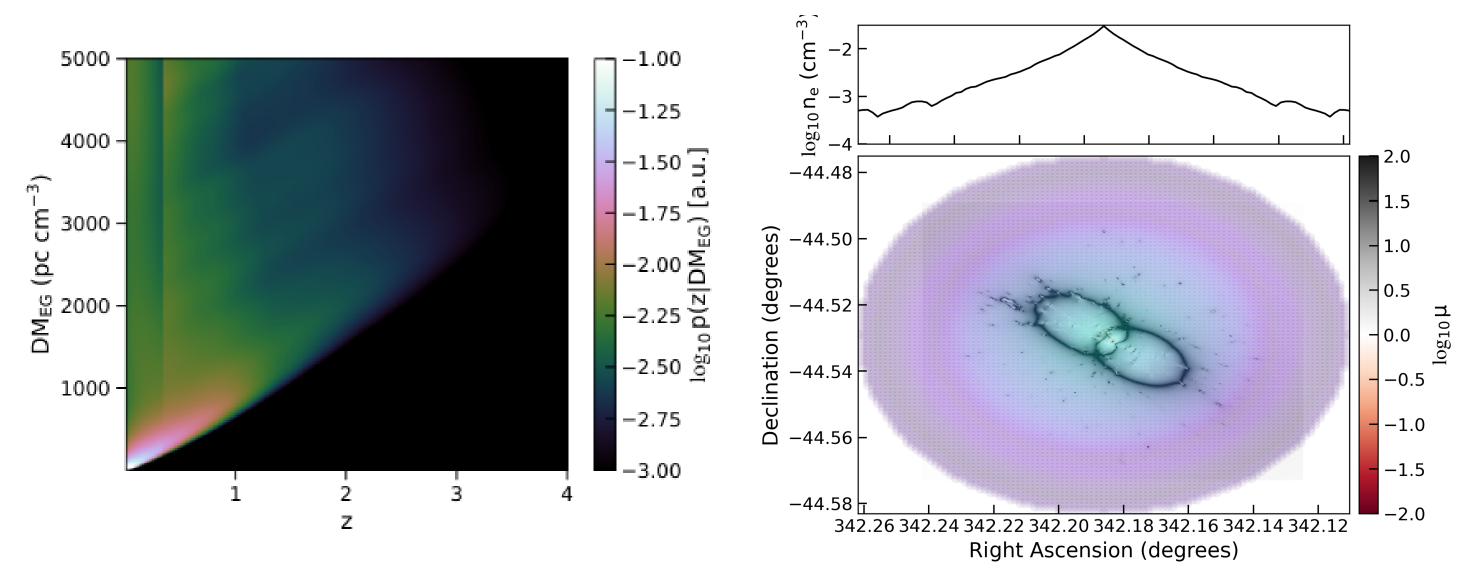}
    \includegraphics[width=0.6\linewidth]{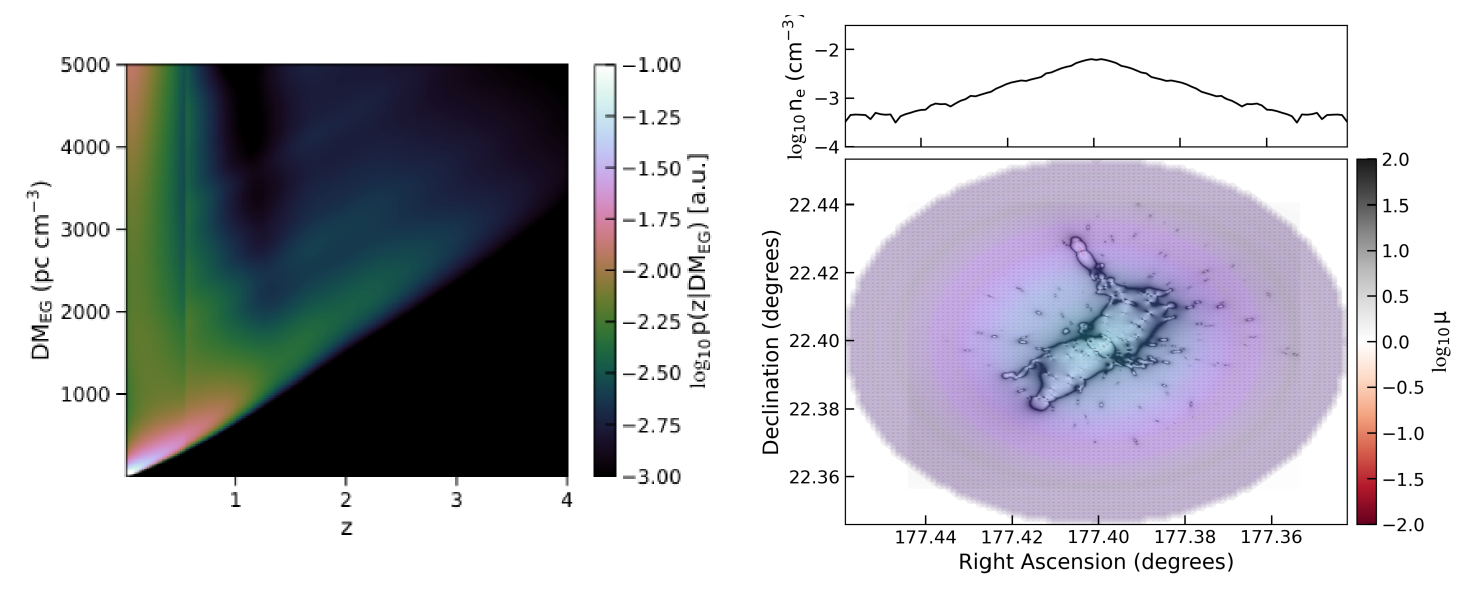}
    \includegraphics[width=0.6\linewidth]{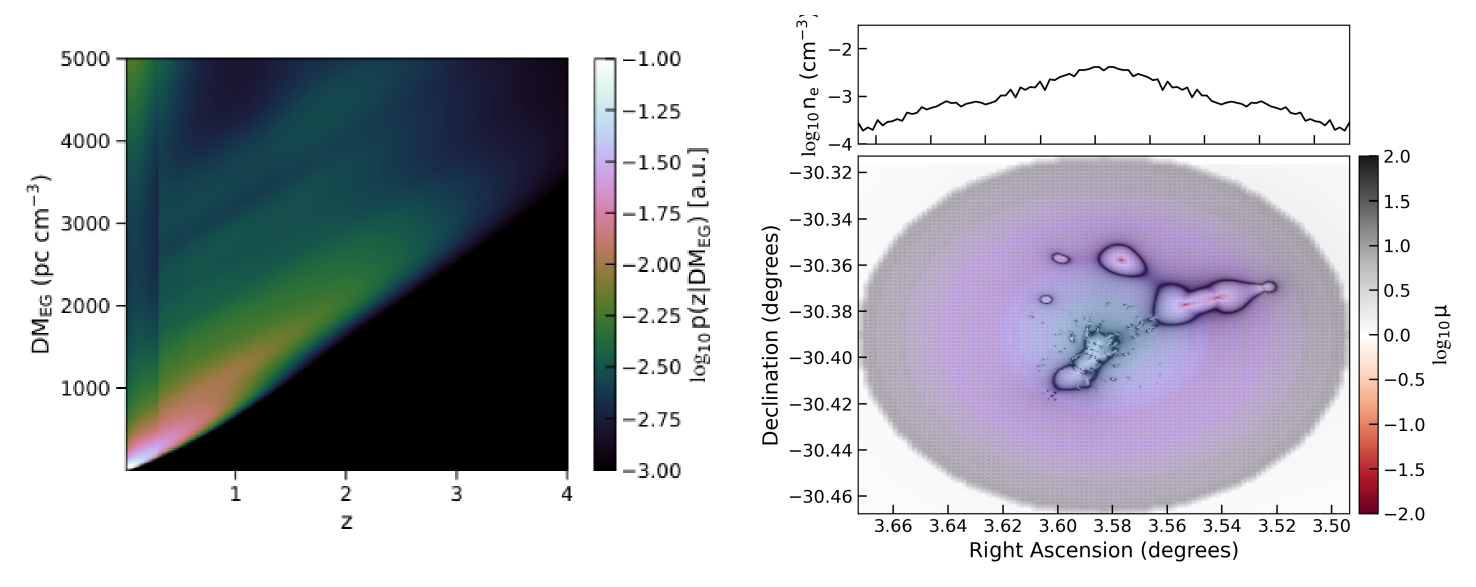}
    \includegraphics[width=0.6\linewidth]{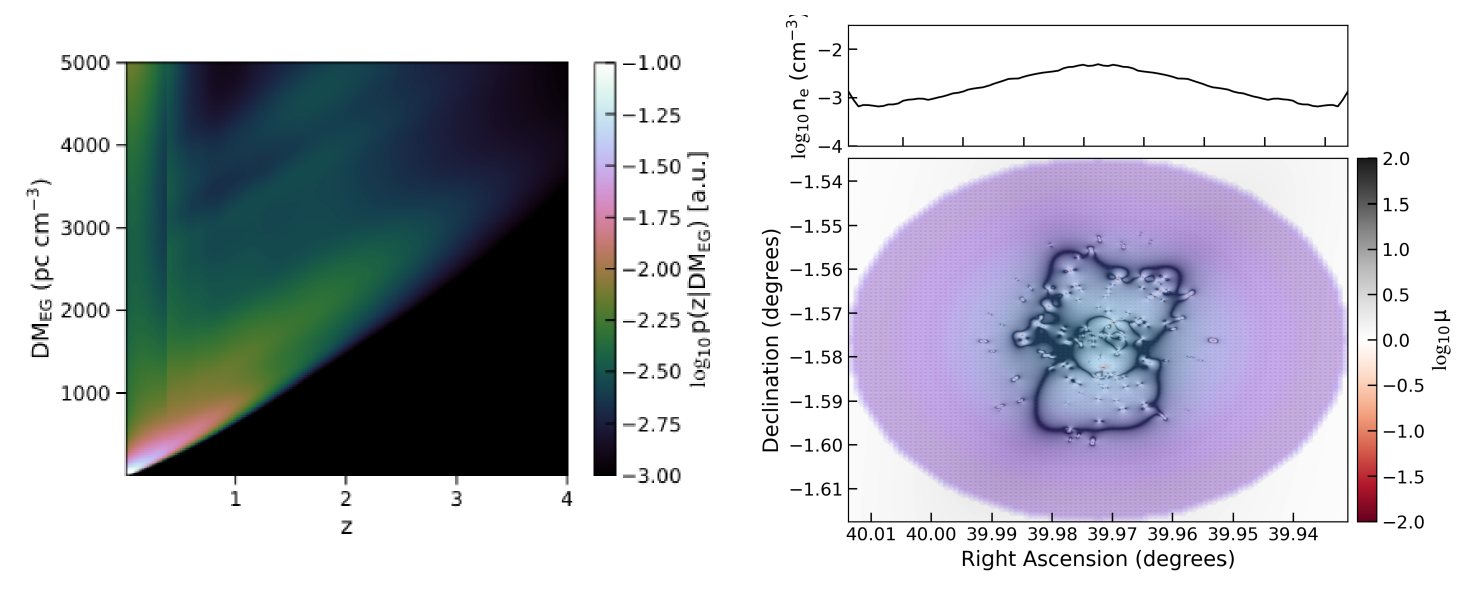}
    \caption{z--DM and cluster morphologies for the high mass Frontier fiedls clusters used in our sample, in descending mass order, MACS J0717.5+3745, AS1063, MACS J1149.5+2223, Abell 2744 and Abell 370}
    \label{fig:highmasszdmmorph}
\end{figure*}

\begin{figure*}
    \centering
    \includegraphics[width=0.6\linewidth]{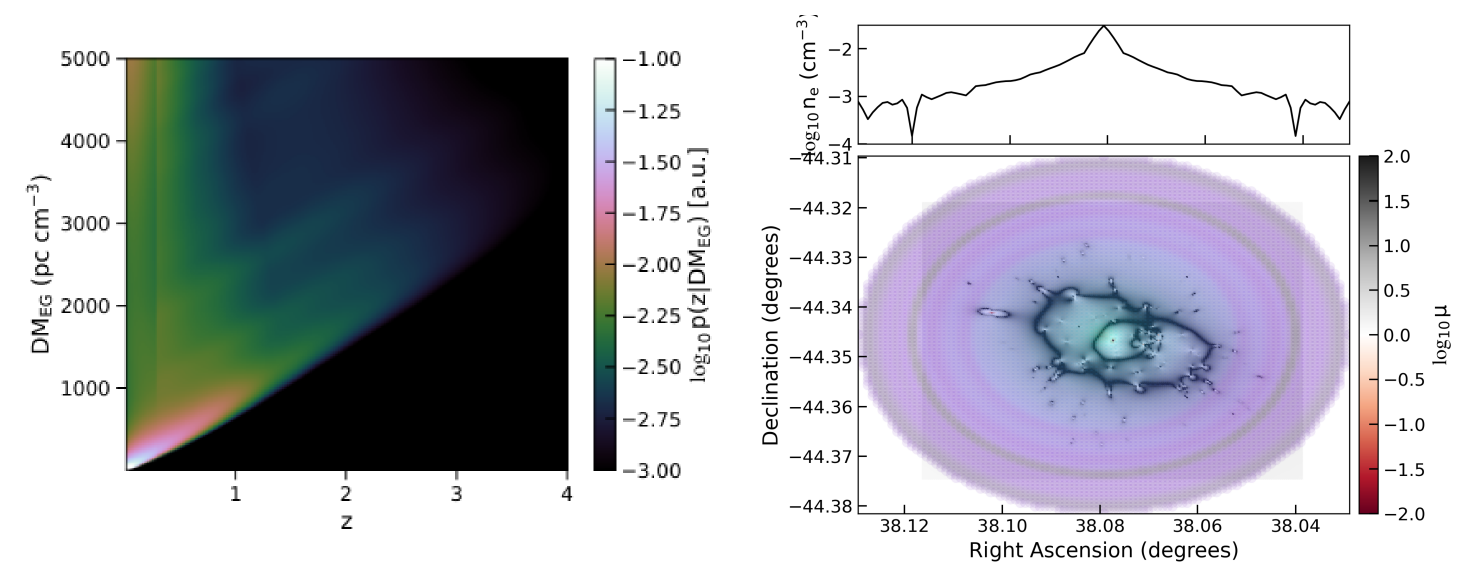}
    \includegraphics[width=0.6\linewidth]{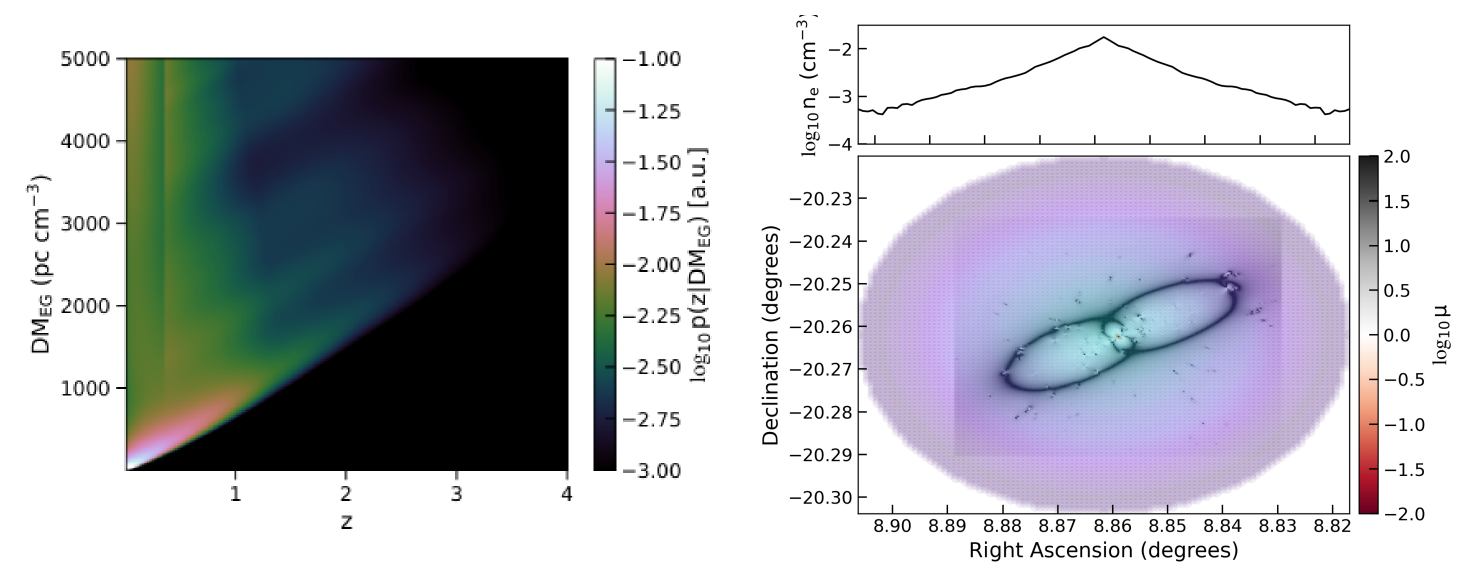}
    \includegraphics[width=0.6\linewidth]{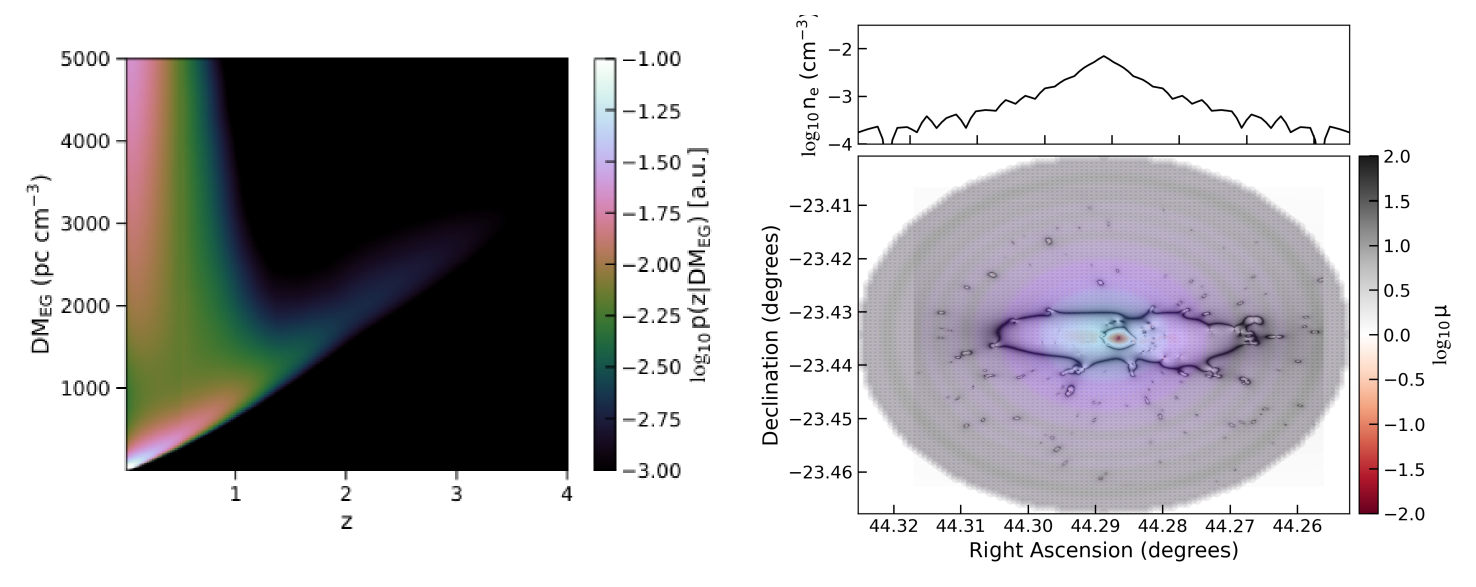}
    \includegraphics[width=0.6\linewidth]{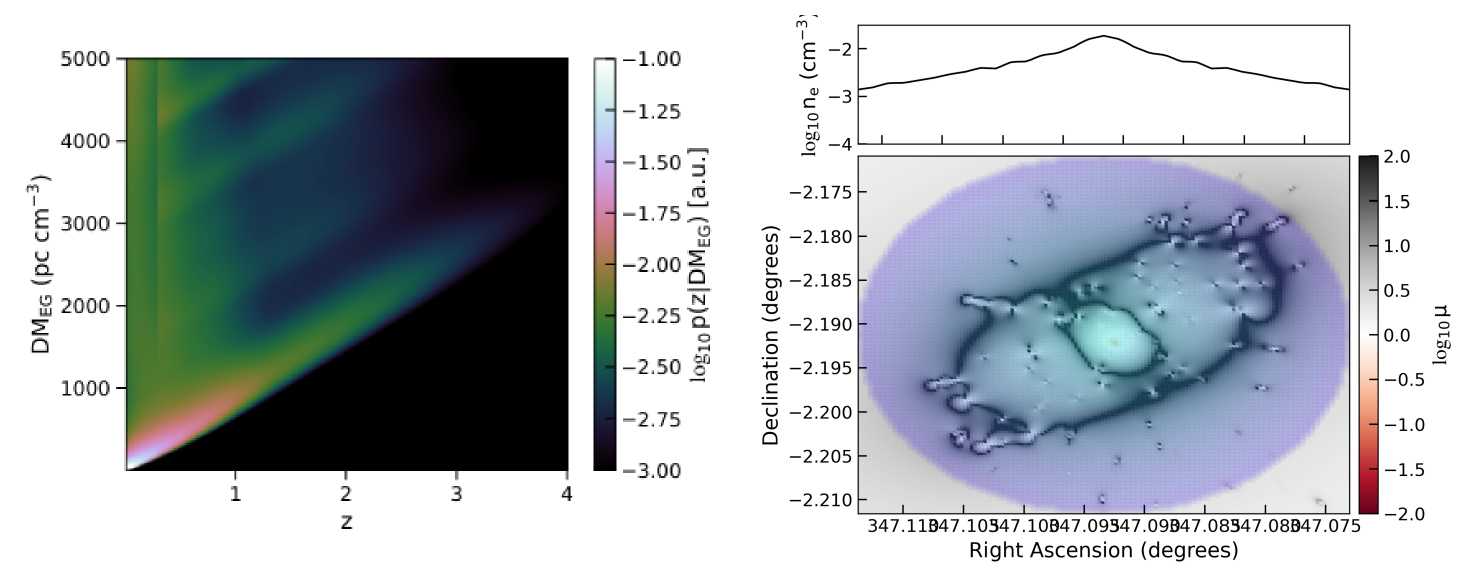}
    \includegraphics[width=0.6\linewidth]{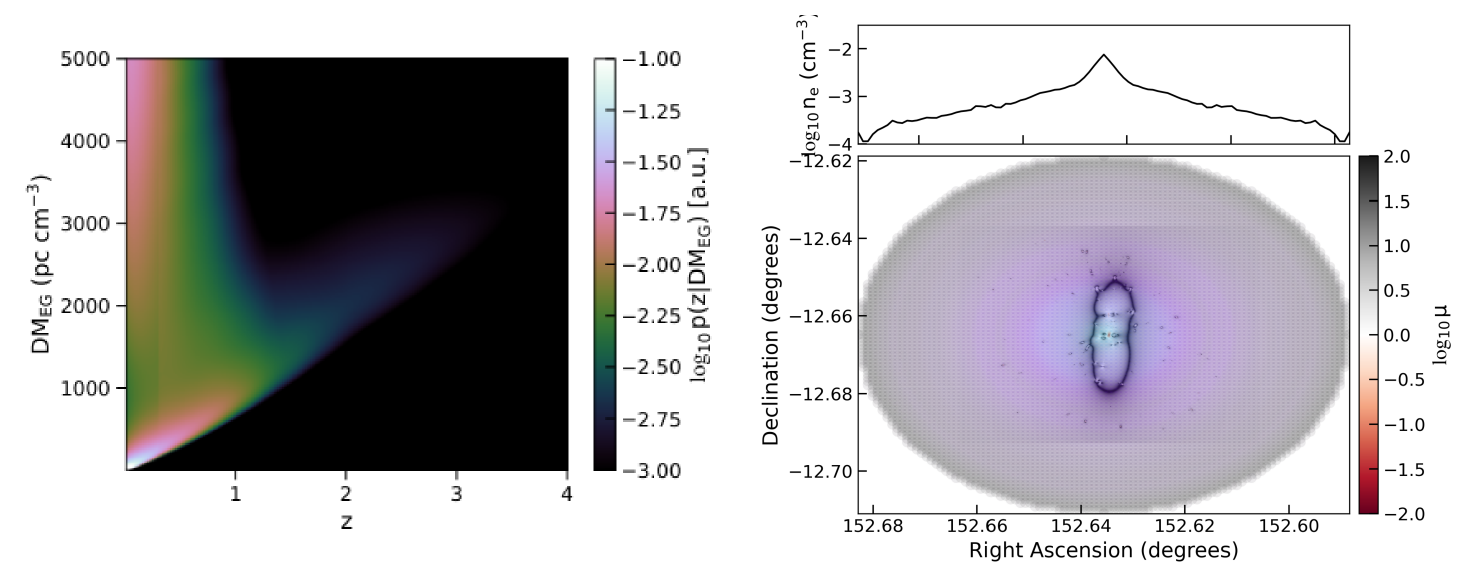}
    \caption{z--DM and cluster morphologies for the low mass RELICS clusters used in our sample, in descending mass order, RXC J0232.2-4420, MACS J0035.4-2015, MACS J0257.1-2325, Abell 2537 and MS 1008.1-1224.}
    \label{fig:lowmasszdmmorph}
\end{figure*}

\section{Lens Model Uncertainty}\label{app:magUncertainty}
\begin{figure}
    \centering
    \includegraphics[width=0.8\linewidth]{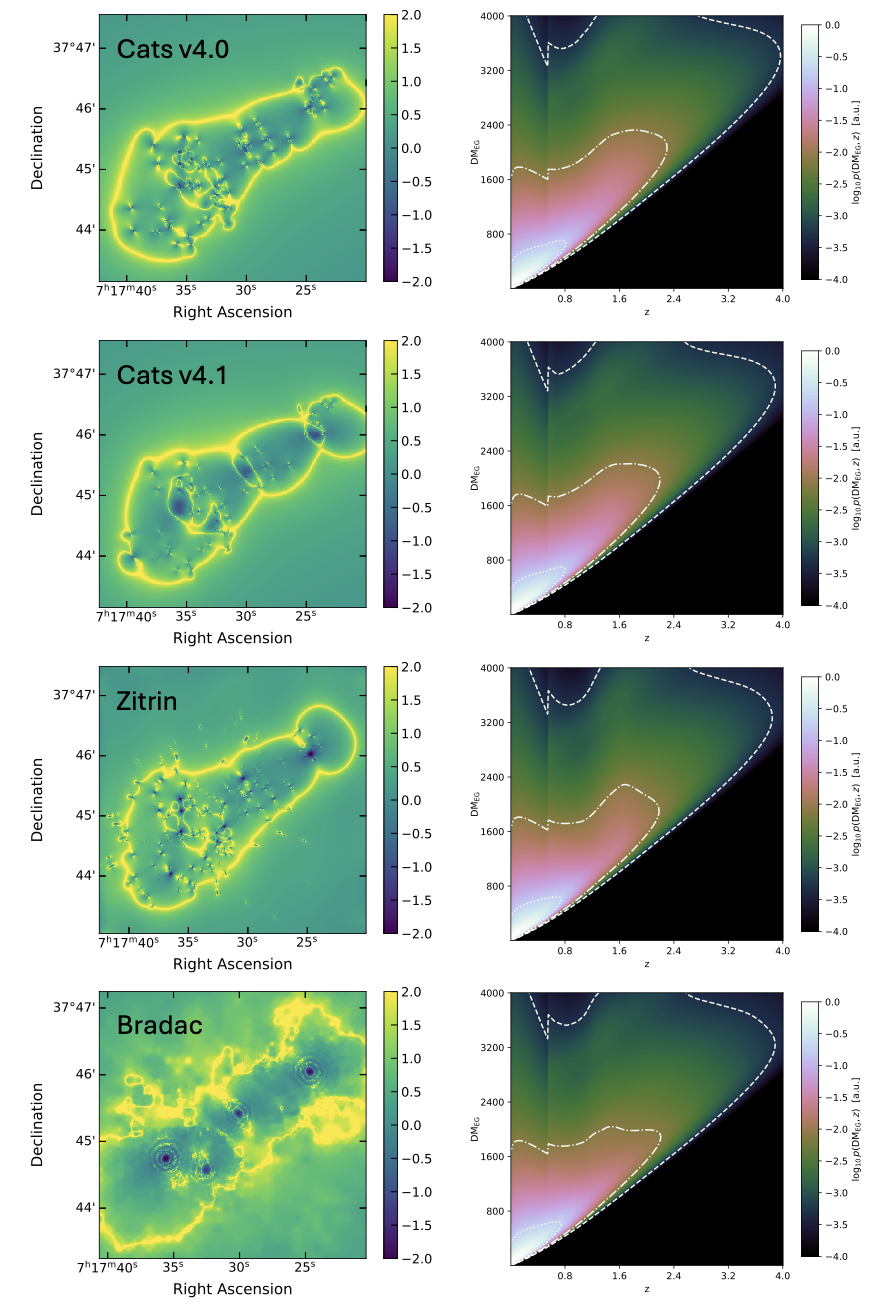}
    \caption{Effect of magnification model uncertainty on observed FRB rates. \textit{Left:} Varying magnification models for MACS J0717.5+3745. \textit{Right:} z--DM distributions corresponding to each lens models.}
    \label{fig:magUncertainty}
\end{figure}

In general, uncertainties in the cluster mass distribution lead to several available lens models for a given cluster. To determine the impact of these model uncertainties on our rate expectations we calculate rates for several models of MACS J0717.5+3745, as shown in Fig. \ref{fig:magUncertainty} for the CHIME telescope. For significant morphological changes in the morphology of the magnification distribution we see only slight changes in the corresponding z--DM distribution of FRBs from the cluster line of sight, and order $10\%$ changes in the absolute rate of detected bursts. As such we conclude that the impact of lens model uncertainties does not substantially change the conclusions we draw in this work.

\label{lastpage}
\end{document}